\documentclass[prl,twocolumn,superscriptaddress,showpacs,amsmath,amssymb]{revtex4-1}
\usepackage{graphicx}
\usepackage{amsmath}
\usepackage{amssymb}
\usepackage{float}
\usepackage{color}

\usepackage{cancel,xcolor}
\usepackage{indentfirst}
\usepackage{CJKulem}
\usepackage{soul,xcolor}
\setstcolor{red}
\usepackage{booktabs}
\usepackage{ulem}
\usepackage{cancel}
\usepackage[pdfpagemode=UseNone,pdfstartview=FitH,colorlinks=true,linkcolor=blue,urlcolor=blue,anchorcolor=blue,citecolor=blue]{hyperref}

\newcommand{\beq}{\begin{equation}}
\newcommand{\eeq}{\end{equation}}
\newcommand{\beqa}{\begin{eqnarray}}
\newcommand{\eeqa}{\end{eqnarray}}

\def\Aa1#1{\textcolor{blue}{#1}}

\def\prl#1{{ Phys.\ Rev.\ Lett.} {\bf#1}}

\def\b0{{\bf{0}}}

\begin{document}
\title{ Scattering description of edge states  in  Aharonov-Bohm triangle chains}


\author{Zhi-Hai\!  Liu }
\email{liuzh@baqis.ac.cn}
\affiliation{Beijing Academy of Quantum Information Sciences, Beijing 100193, China }

\author{O. Entin-Wohlman}
\email{orawohlman@gmail.com}
\affiliation{School of Physics and Astronomy, Tel Aviv University, Tel Aviv 6997801, Israel}

\author{A. Aharony}
\affiliation{School of Physics and Astronomy, Tel Aviv University, Tel Aviv 6997801, Israel}

\author{J. Q. You}
\affiliation{School of Physics, Zhejiang University, Hangzhou 310027, China}

 \author{H. Q. Xu}
 \affiliation{Beijing Key Laboratory of Quantum Devices, Key Laboratory for the Physics and Chemistry of Nanodevices, and School of Electronics, Peking University, Beijing 100871, China}
 \affiliation{Beijing Academy of Quantum Information Sciences, Beijing 100193, China }

\begin{abstract}

Scattering theory has been suggested as a convenient method to identify topological phases of matter,  in particular of disordered systems for which the Bloch band-theory approach is inapplicable. Here we examine this idea, employing as a benchmark a one-dimensional triangle chain whose versatility yields a scattering matrix that ``flows" in parameter space among several members of the topology classification scheme. Our results show that the reflection amplitudes (from both ends of a sufficiently long chain) do indicate the appearance of edge states in {\it all} (topological and non-topological) cases. For the topological cases, the transmission has a peak at the topological phase transition, which happens at the Fermi energy. A peak still exists as one moves into the non-topological `trivial' regions,
in which another transmission peak may occur at nonzero energy, at which a relevant edge state appears in the isolated chain. For finite chains, the peak in the transmission strongly depends on their coupling of the leads, and {\it not} on the phase transition of the isolated chain.  In any case, {\it the appearance of a peak in the transmission is not sufficient to conclude that the system undergoes a topological phase transition.}

 \end{abstract}


\date{\today}
\maketitle

\noindent{\color{blue}\bf{Introduction.---}}Building on the bulk-boundary correspondence,   topological phases of matter are characterized by the emergence of edge or surface states at the Fermi energy~\cite{Hasan2010,Qi2011,Bansil2016}, between the system and the vacuum.  Bound states in a finite system affect its scattering properties, suggesting \cite{Fulga2011} the use of the scattering amplitudes to detect topological phases. This possibility was demonstrated by Fulga {\it et al.}, who showed that the reflection matrix of a semi-infinite low-dimensional wire obeying a highly-symmetric Hamiltonian reflects its topological quantum invariant~\cite{Fulga2011,Fulga2012}.
Scattering theory, as opposed to the Bloch band-structure formalism, operates in real space and thus is, in particular, suitable for studying topological properties of disordered or inhomogeneous systems~
\cite{Sbierski2014}, 
inaccessible in the band-structure picture~\cite{Bansil2016}. Conversely, it necessarily connects a finite system to its (non-topological) surroundings \cite{Bissonnette2023}.
Our aim is to examine the effectiveness of this method for predicting and detecting topological features.


In the {\it elastic} scattering-matrix formalism, the system (i.e., a finite one-dimensional wire in our case) is attached on both sides to two semi-infinite leads, usually described by tight-binding chains.
The scattering matrix of such a structure depends naturally on the parameters of these leads,  which fix the scattering energy,
and on their coupling with the system. These points were largely avoided in previous studies, primarily because they set the scattering energy at the Fermi level~\cite{Fulga2011},
adopted an approximate model-independent form of the scattering matrix \cite{Mahaux1968}, and focused on the reflection amplitude of a semi-infinite system.
Another complication arises from the dependence of the scattering amplitudes on the wire's length. For instance, due to the finite-size effect,    the threshold for the emergence of nontrivial edge states in a {\it finite} Su-Schrieffer-Heeger (SSH) chain~\cite{Delplace2011,Bissonnette2023} deviates from the criterion for a topological phase-transition in the model
~\cite{Su1979,Atala2013}.    The influence of this ubiquitous size effect on topological features and on the scattering amplitudes has not yet attracted much attention, though the scattering formalism seems to be the best tool for exploring the topology of disordered systems.


In addition to the {\it topologically-nontrivial symmetry classes} \cite{Altland1997,Ryu2010}, it has been found recently
 that systems lacking quantized topological invariants, thus belonging to topologically {\it trivial symmetry classes}, may exhibit intriguing phenomena. For instance,  unexpected bound states, observed experimentally
 in photonic lattices \cite{Aravena2022}, Rydberg-atom synthetic dimensions \cite{Kanungo2022},  atomic Aharonov-Bohm (AB)  cages \cite{Mukherjee2018,Kremer2020,Gou2020,Li2022}, or theoretical proposals, e.g.,  topology in laser systems \cite{Harari2018}. Whether the scattering approach is capable of analyzing the fate of the edge states, as a low-dimensional chain moves in parameter space away from topologically nontrivial symmetry classes to topologically trivial ones, is an open question.  Sorting out these points is the scope of this paper.



\begin{figure}
\includegraphics[width=0.42\textwidth]{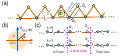}
\caption{(color online)  (a) A  triangle chain,
hosting two sites $a$ and $b$     in the unit cell, with   $j^{}_{1,2}$ and  $j^{}_{3}$ being the in-cell and out-of-cell tunneling amplitudes.   Each triangle is threaded by an  Aharonov-Bohm flux $\Phi $ in units of the flux quantum.  (b) The possible symmetry classes of the triangle chain in the  $(|j^{}_{3}|,\Phi)-$ plane,   with the  ``BDI''
class at the origin, the  ``AI'' (``D'') class on the horizontal (vertical)  axis, and the ``A'' class occupying the entire plane  (except the axes). (c) Two possible  configurations of coupling a finite chain of $N$ cells to two semi-infinite leads: in both configurations, the left lead is coupled  to  $a^{}_{1}$-site,
$\lambda=0$ (1)  describe  coupling  the right lead to   a $b^{}_{N}$ ($a^{}_{N}$)-site.   The tunneling amplitudes on the leads are $J^{}_{0}$, and $J^{}_{\rm c}$ is the coupling amplitude.  }
\label{Fig1}
\end{figure}


As a benchmark, we study a one-dimensional chain whose versatility yields a scattering matrix that ``flows" in parameter space among several members of the topology classification scheme \cite{Altland1997,Ryu2010},   including the ``BDI" class, which contains the SSH model \cite{Su1979}.
The latter is one of the
simplest models out of which emerge interesting topological properties~\cite{Agrawal2023,Jean2017}.  Another route of research focuses on   open SSH chains coupled to leads \cite{Bissonnette2023,Ostahie2021, Garmon2021}.

As expected, the finiteness of the chain, combined with details of its coupling with the semi-infinite leads,  affect the scattering amplitudes and thus may hinder their use as indicators of topological properties characterizing the isolated chain. For instance, focusing on the reflection amplitude at the Fermi energy of a topological chain \cite{Fulga2011,Fulga2012}, we find that though the size effect may be removed by making the chain semi-infinite (and coupling it to a single lead),  the {\it phase} of the reflection amplitude does depend on the endpoint at which it is monitored.
Interestingly, the reflection amplitude of a semi-infinite chain belonging to a topologically trivial symmetry class
can  attain the `nontrivial' value $-1$  but at an energy different from the Fermi energy. Moreover, an edge state appears in a long, isolated chain at that energy. Such edge states of topologically trivial chains can be robust against disorder.


Our results show that the reflection amplitudes (from both endpoints of a finite long chain) do indicate the appearance of edge states in {\it all} (topological and non-topological) cases. For the topological cases, the transmission has a peak at the topological phase transition, which happens at the Fermi energy $E=0$, {\it even for finite long chains}. This peak decays as the chain moves into the non-topological trivial regions,
in which a  transmission peak may also occur at nonzero energy.  In any case, {\it the appearance of a peak in the transmission is not sufficient to conclude that the system undergoes a topological phase transition}. Note that a standard scattering experiment measures the reflection magnitude $|r|^2$, not its amplitude $r$. However, even if the sign of $r$ could be measured, it does {\it not} identify the topological phase.


\vspace{3mm}
\noindent{\color{blue}\bf{The chain.---}} The chain we analyzed [Fig. \ref{Fig1}(a)] is built of identical triangles,  each penetrated by an AB magnetic flux
$\Phi$. The vertices are two different sites, $a$ and $b$, connected by staggered tunneling amplitudes, $j^{}_{1}$ and $j^{}_{2}$.  Nearest-neighbor $a$ sites are connected by
 $j^{}_{3}$.   The     Hamiltonian of the  triangle chain  is
 \begin{align}
 {\cal H}^{}_{\Delta}&= - \sum^{}_{n}\Big(j^{}_{1}a^{\dagger}_{n}e^{i\phi^{}_{1}}_{}b^{}_{n} +j^{}_{2}b^{\dagger}_{n}e^{i\phi^{}_{2}}_{}a^{}_{n+1}\nonumber\\
&+j^{}_{3}a^{\dagger}_{n}e^{i\phi^{}_{3}}_{}a^{}_{n+1}+{\rm h.c.}\Big)\ ,
\end{align}
  where $a^{\dagger}_{n}$  ($b^{\dagger}_{n}$) is the creation operator of a spinless electron residing on the $a$ ($b$) site of the  $n$-th cell and  $\phi^{}_{i=1,2,3}$   are the partial AB   fluxes (in units of  $\Phi^{}_{0}=h/e$), accumulated on each bond, with  $\Phi=\phi^{}_{1}+\phi^{}_{2}-\phi^{}_{3}$.     Contingent on the values of $j^{}_{3}$ and $\Phi$,  this chain ``flows'' among four symmetry classes: For $j^{}_{3}=0$ the AB flux is irrelevant and the chain becomes the SSH wire \cite{Su1979}, whose Hamiltonian obeys time-reversal, particle-hole, and chiral symmetries and thus belongs to the ``BDI" class.  For a nonzero  $j^{}_{3}$ and $\Phi=\pm  \pi/2$, the Hamiltonian is particle-hole symmetric (details in Ref. \cite{Supplementary}), placing the chain in the ``D" class. For other parameters the system belongs to the topologically-trivial classes ``A" and ``AI", see  Fig.~\ref{Fig1}(b).


\vspace{3mm}
\noindent{\color{blue}\bf{Scattering Matrix.---}} The scattering formalism requires attaching semi-infinite leads
on both sides of a finite chain. Modeling those by identical one-dimensional tight-binding  Hamiltonians (with zero on-site energies and tunneling amplitudes $J^{}_0$), and denoting both tunnel couplings with the chain by $J^{}_{\rm c}$, one encounters two possible edge configurations, see  Fig.~\ref{Fig1}(c): in the $\lambda=0$ configuration the chain comprises an integer number $N$ of unit cells,  for $\lambda=1$ it does not, the rightmost site being $a^{}_{N}$.


 The scattering matrix ${\cal S}$ relates  the amplitudes of the incoming  plane-waves from the left (right),
 $A^{}_{L}$ ($A^{}_{R}$),  with those of the outgoing waves,  $B^{}_{L (R)}$, whose energy $E$ is related to the (dimensionless) wave vector $\kappa$,
\begin{align}
\left[\begin{array}{c}
 B^{}_{L}\\
 B^{}_{R}
 \end{array}
 \right]={\cal S }^{}_{\lambda} \left[\begin{array}{c}
 A^{}_{L}\\
 A^{}_{R}
\end{array}\right]\ ,\ \
E^{}_{}=-2J^{}_{0}\cos(\kappa)\ .
\label{En}
\end{align}
Denoting by $\Psi^{}_{a /b } (n)$,  ($n=1,2 ,...,N$),  the wave functions at   the   $a$/$b$  sites  of  the $n$-th  cell,  continuity at the endpoints implies
\begin{align}
\frac{J^{}_{\rm c}}{J^{}_{0}}\left[
\begin{array}{c}
\Psi^{}_{a}(1)\\
\Psi^{}_{\chi}(N)
\end{array}
\right]=e^{i\kappa}  \left[
\begin{array}{c}
A^{}_{L}\\
A^{}_{R}
\end{array}
\right]+e^{-i\kappa}  \left[
\begin{array}{c}
B^{}_{L}\\
B^{}_{R}
\end{array}
\right]\ , \label{SV2}
\end{align}
where    $\chi= b(a)$ for the $\lambda=0 (1)$ configuration.


Another relation for the wave functions at the two endpoints is obtained by the transfer matrix.
Let $\mathbf{V}^{ }_{n}=\{\Psi^{}_{b} (n),\Psi^{}_{a}(n)\}^{T}_{}$
  be a  vector comprising the wave functions at the two sites of the unit cell,     with $1\leq n\leq N-2$.
  As  these   functions   are eigenfunctions of  ${\cal H}^{}_{\Delta}$, the   transfer matrix  connecting  two    nearest-neighbor  vectors,  $
 \mathbf{V}^{ }_{n+1}={\cal T}(E)\mathbf{V}^{ }_{n}
$,  is
\begin{align}
&{\cal T}(E)= \exp\left(\alpha \hat{\bf{v}}\cdot \boldsymbol{\sigma}-i\zeta\right)\ ,
\label{TMA}
\end{align}
with $ \cosh(\alpha)=(E^{2}_{}-j^{2}_{1}-j^{2}_{2})/(2|W|)$
($\boldsymbol{\sigma}$ is the vector of the Pauli matrices).
Here,
$W=j^{}_{1}j^{}_{2}-Ej^{}_{3}e^{-i\Phi}_{}\equiv |W|\exp(i\beta)$,  $\zeta=\Phi+\phi^{}_{3}+\beta$, and  $\hat{\bf{v}}$ is a parameters-dependent  two-dimensional vector, given in Ref. {\cite{Supplementary}.


As usual, the Bloch spectrum of a  chain obeying periodic boundary conditions is given by $ {\cal T}^{N}_{} \mathbf{V}^{}_{n}=\mathbf{V}^{}_{n}$.  In that case, $\alpha$  is purely {\it imaginary}  for a bulk state, while when $E$ is located in the band gap $\exp[\alpha(E)]$ is real and smaller than $-1$ ~\cite{Supplementary}.
In an open chain, for which translational symmetry is broken around  the  chain-lead junctions,  the  connection between $\mathbf{V}^{ }_{N-1}$ and $\mathbf{V}^{ }_{N}$
depends on the specific endpoints [Fig. \ref{Fig1}(c)].   We find
\begin{align}
{\cal F}^{}_{\lambda}(E)
\left[\begin{array}{c}
\Psi^{}_{a}(1)\\
\Psi^{}_{\chi}(N)
\end{array}\right] =J^{}_{\rm c}\left[\begin{array}{c}
A^{}_{L}+B^{}_{L}\\
A^{}_{R}+B^{}_{R}
\end{array}\right]\ .
\label{SV1}
\end{align}
The  $2\times 2$  matrix ${\cal F}^{}_{\lambda}(E)$ is given in Ref. \cite{Supplementary}.  When the  chain is isolated from the leads, i.e.,  $J^{}_{\rm c}=0$, its eigen energies are given by     $\det  [{\cal F}^{}_{\lambda}(E)]=0$. Otherwise,
\begin{align}
 {\cal S }^{}_{\lambda}&\equiv \left[\begin{array}{cc}
  r^{}_{\lambda,l}& t^{ }_{\lambda,r}\\
  t^{}_{\lambda,l} & r^{ }_{\lambda,r}
  \end{array}\right]\nonumber\\
  &=
- 1 -2i\sin(\kappa)  \Big [e^{-i\kappa}_{}-\Lambda{\cal F}^{-1}_{\lambda}\Big ]^{-1}_{}\ ,
\label{SCA}
\end{align}
where $\Lambda=(J^{2}_{\rm c}/J^{}_{0})$,
$r^{}_{\lambda,l}$ ($r^{ }_{\lambda,r}$)  and $t^{}_{\lambda,l}$ ($t^{ }_{\lambda,r}$)
are the reflection and transmission amplitudes  from the left (right) in the $\lambda$ configuration, and the unitarity
 of  ${\cal S}^{}_{\lambda}$ is ensured by  ${\cal F}^{}_{\lambda}={\cal F}^{\dagger}_{\lambda}$.

\begin{figure}
\includegraphics[width=0.42\textwidth]{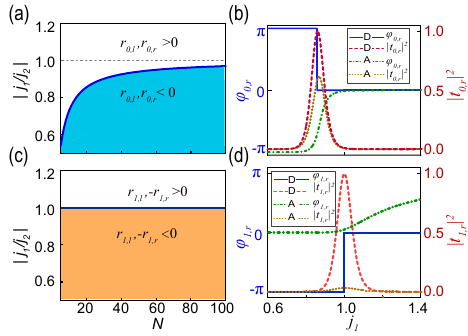}
\caption{(color online)   (a) The distribution of the sign of the two-sided reflection amplitudes off a finite chain in the $\lambda= 0$ configuration.
The (blue) thick curve
is the threshold for the sign change.  (b)  The corresponding reflection phases   $\varphi^{}_{0,r} $    and   the transmission $|t^{}_{0,r}|^{2}$  
of a finite chain of $N=20$ cells in the ``A'' ($\Phi=0.48$) and ``D'' classes. (c) The same as  (a), for a chain in the $\lambda=1$ configuration. The horizontal line 
marks the threshold for the sign change. (d)     $\varphi^{}_{1,r} $    and  $|t^{}_{1,r}|^{2}$  
for the finite chain with $\lambda= 1$.  Here, the scattering energy is fixed at the Fermi level ($E=0$)  and the tunneling amplitudes, in units of $j^{}_{2}$,   are  $j^{}_{3}=0.2$,  $J^{}_{\rm c}=0.3$, and $J^{}_{\rm 0}=2.0$.  }
 \label{Fig2}
\end{figure}


\vspace{3mm}
\noindent{\color{blue}\bf{Size effects.---}}      We begin this discussion with the $\lambda=0$ configuration.  The scattering amplitudes  on both sides  of a finite chain,  at the Fermi energy ($E=0$),  are
 \begin{align}
r^{}_{0,l/r}
 =-\frac{ \Lambda^{2}_{}e^{(N-1)\alpha}_{} \pm2i\xi\Lambda S^{}_{N-1}  +|W| e^{-N\alpha }_{} }{\Lambda^{2}_{}e^{(N-1)\alpha}_{}  -2i\xi\Lambda S^{}_{N-1} - |W| e^{-N\alpha }_{}}\ , \label{r0w}
\end{align}
where $\xi=2j^{}_{1}j^{}_{2}j^{}_{3}\cos\Phi/
 (j^{2}_{1}-j^{2}_{2}) $,  $S^{}_{n} =\sinh[n\alpha]$,     $\exp[\alpha]=-|j^{}_{2}/j^{}_{1}|$ for $E=0$.
As $\xi=0$ for the most symmetric classes    ``BDI'' ($j^{}_{3}=0$) and   ``D'' ($\Phi=\pi/2$),  their reflection amplitudes of both sides  are equal,
\begin{align}
r^{}_{0,l}=r ^{}_{0,r}=\Big( \left|\frac{j^{}_{1}}{j^{}_{2}}\right|^{2N}_{}   - \frac{\Lambda^{2}}{j^{2}_{2}} \Big) \Big(  \left|\frac{j^{}_{1}}{j^{}_{2}}\right|^{2N}_{}   + \frac{\Lambda^{2}}{j^{2}_{2}} \Big)^{-1}_{}
\ .
\end{align}
The reflection amplitude is  negative  for
 $|j^{}_{1}/j^{}_{2}|<\gamma^{}_{N}$ with $\gamma^{}_{N}=|\Lambda/j^{}_{2}|^{1/N}$ being the parameter indicating the finite-size effect, and is positive otherwise, see Fig.~\ref{Fig2}(a). At  $|j^{}_{1}/j^{}_{2}|=\gamma^{}_{N}$, the reflection amplitude changes from $-1$ to $+1$, and the transmission has a peak.
 This is distinct from the size effect on the threshold for the emergence of edge states in the isolated SSH chain, i.e., $|j^{}_{1}/j^{}_{2}|<N/(N+1)$ \cite{Delplace2011}, which results from the spectrum of the isolated chain. 
 Instead,  a transmission peak appears for a finite chain for  $|j^{}_{1}/j^{}_{2}|=\gamma^{}_{N}$, where the reflection passes through zero, as seen in Fig.~\ref{Fig2}(b).   That peak coincides with that of Ref. \onlinecite{Delplace2011} only for the semi-infinite chain, when $\gamma^{}_N=1$. The appearance of such a  peak in that limit has been proposed as a signal of the topological phase-transition  (TPT)~\cite{Fulga2011}. Its dependence on the coupling with the leads for finite chains casts doubts on this identification.

As  $\xi\neq 0$ for the two low-symmetry classes  ``AI'' and ``A'',  the reflection amplitudes of both sides in  Eq.~(\ref{r0w})  are complex and not equal to one another.  The reflection phases $\varphi^{}_{0,l}\equiv \arg[r^{}_{0,l}]$ and $\varphi^{ }_{0,r}\equiv \arg[r^{ }_{0,r}]$ vary smoothly with  $|j^{}_{1}/j^{}_{2}|$, and the reflection amplitudes do not cross zero around $|j^{}_{1}/j^{}_{2}|=1$,  excluding the emergence of a sharp transmission peak of unit magnitude. Interestingly, such a (smaller) peak still appears near the topological axes in Fig. \ref{Fig1}(b), decaying as one moves away from these axes. This peak does not imply a TPT!

We now turn to the $\lambda=1$ configuration.
In this case, the size effect can be removed.  For $\lambda=1$ the reflection amplitudes   at the Fermi energy are
 \begin{align}
r^{}_{1,l/r } = \frac{\xi \mp i\Lambda}   { \xi  +i\Lambda \coth^{}_{}[(N-1)\alpha]}\ .
\label{rlr}
\end{align}
Thus $r^{}_{1,l}=-r^{}_{1,r}= - \tanh^{}_{}[(N-1)\alpha]$
for the ``BDI'' and ``D'' classes (where $\xi=0$). Since at the Fermi energy $\exp[\alpha]_{}=-|j^{}_{1}/j^{}_{2}|$,  the left (right) reflection is negative (positive)  for $|j^{}_{1}/j^{}_{2}|<1$, changing the sign of  $r^{}_{1,l (r)}$ in accordance with the criterion for the TPT.
 However, the different signs of $r^{}_{1,l (r)}$
may lead to conflicting predictions of the criterion for the TPT as presented in Ref.~\onlinecite{Fulga2011}. For instance,  when $|j^{}_{1}/j^{}_{2}|<1$  the presence of   $r^{}_{1,l}=-1$  on the left side indicates a topologically-nontrivial phase, but   $r^{}_{1,r}=1$  on the other side corresponds to a  trivial phase.
 This contradiction can be traced to the lack of a $b$ site at the rightmost ($N$th) cell, making the chain `incomplete'. Instead of a simple two-sublattice model,   the finite chain becomes a `hybrid' system, see details in Ref.~\cite{Supplementary}. Thus, the one-sided reflection does not fully determine the (finite)  system topology.
In any case, the reflection amplitudes calculated by the scattering formalism are not measured directly, and therefore connecting their signs to topology is not very useful. Peaks of the transmission are predicted in many cases, but their relation to the TPTs for finite chains remains questionable.


\vspace{3mm}
\noindent{\color{blue}\bf{Edge states in the trivial symmetry classes.---}} As seen in Eq. (\ref{rlr}), the reflection amplitudes at the Fermi energy of the trivial ``AI'' and ``A'' classes are complex, thus cannot exhibit a jump from $-1$ to $+1$  upon changing
$|j^{}_{1}/j^{}_{2}|$.
However, examining the variations of the reflection phases in Figs.~\ref{Fig3}(a,b)
one notes that a negative reflection amplitude can
appear at certain points in the $j^{}_{1}-E$ plane for these classes, depending on the endpoints' configuration
 \cite{Supplementary} (Recall that in the topological ``D" class this occurs at zero energy.) Figures \ref{Fig3}(c,d)
show that edge states appear in accordance with the negative reflection amplitudes.


The general expression for the   reflection amplitude reads [see Eq.~(\ref{SCA}) and Ref. \cite{Supplementary}]
\begin{align}
r^{}_{\lambda,l/r} =-1-\frac{2i(\det[{\cal F}^{}_{\lambda}]e^{-i\kappa}_{}-\Lambda {\cal F}^{}_{\lambda,11/22})\sin \kappa}{\det[{\cal F}^{}_{\lambda}]e^{-2i\kappa}_{}-\Lambda {\rm tr}[{\cal F}^{}_{\lambda}]e^{-i\kappa}_{}+\Lambda^{2}_{}}\ . \label{rf}
\end{align}
 When the scattering energy resonates with one of the electronic states of the isolated chain, implying that $\det[{\cal F}_{\lambda}(E)]=0$,  a   negative reflection on the left/right endpoint appears only if   ${\cal F}^{}_{\lambda,11/22}=0$.  However, in a finite chain the two conditions are not compatible:   for ${\cal F}^{}_{\lambda, nn}=0$,    $\det[{\cal F}^{}_{\lambda}]= -{\cal F}^{}_{\lambda,12}{\cal F}^{}_{\lambda,21}\propto \exp[-2N\alpha ]$   which generally differs from zero,  unless  $E$ is located in the energy gap  ($\exp[\alpha(E)]< -1$) and $N\rightarrow \infty$. Thus,  a left/right edge state in the band gap of a sufficiently long isolated chain can appear together with a negative reflection amplitude from that side, provided that  ${\cal F}^{}_{\lambda,11/22 }=0$.


\begin{figure}
\includegraphics[width=0.46\textwidth]{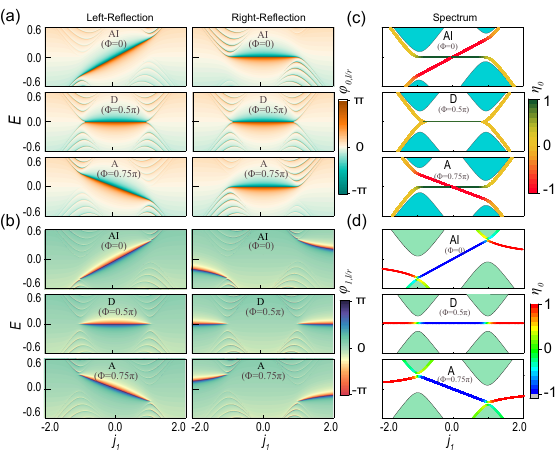}
\caption{(color online) Comparison of features of the  ``AI", ``D", and ``A"  classes (the corresponding Aharonov-Bohm fluxes are marked on the panels)  in the $\lambda=0$ [panels (a) and (c)] and $\lambda=1$  [panels (b) and (d)] configurations.
(a,b) The phases' distribution of the reflection amplitudes off the two endpoints of a finite chain in the $j^{}_{1}-E$ plane.  (c,d)  The energy spectra of isolated chains as functions of  $j^{}_{1}$.   The sidebar indicates the extent of the energy states in the gap. The computation uses $N=20$, with tunneling amplitudes as in Fig.~\ref{Fig2}. }
\label{Fig3}
\end{figure}


 \begin{figure}
\includegraphics[width=0.475\textwidth]{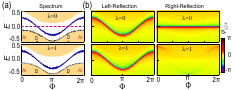}
\caption{(color online)  (a) Energy spectra of a finite disordered isolated chain (parameters as in Fig.~\ref{Fig2}) as functions of  $\Phi$  in the  $\lambda=0,1$ configurations with $j^{}_{1}=0.8j^{}_{2}$.  The left and right edge states are indicated by the solid and dashed lines. (b)  The phases' distribution of the reflection amplitudes off the two endpoints of the finite disordered chain in the two  configurations in the $\Phi-E$ plane.  The disorder is modeled by adding to the tunneling amplitude $j^{}_{1}$ a random term,  $\delta j^{}_{1,n}= 0.3( w^{}_{n}-0.5)$, with $ w^{}_{n}\in [0,1]$ being a  \textit{random} number.}
\label{Fig4}
\end{figure}

Let us examine the conditions required to realize this situation. For the  $\lambda=0$ configuration,
\begin{align}
{\cal F}^{}_{0,22}=& 0~~~\forall ~ ~\{E=0 ,\ \  j^{2}_{2}>j^{2}_{1}\}\ ,\nonumber\\
{\cal F}^{}_{0,11}=& 0~~~\forall~~ \{E=\varepsilon^{}_{2},\ \ \varepsilon^{2}_{2} +j^{2}_{2} >j^{2}_{1} \}\ ,
\label{CY}
 \end{align}
where $\varepsilon^{}_{2}\equiv\xi(j^{2}_{1}-j^{2}_{2})/(j^{2}_{2}+j^{2}_{3})$.  The vanishing of ${\cal F}^{}_{0,22}$  implies the appearance of a  negative reflection amplitude on the right side [see Fig.~\ref{Fig3}(a)], corresponding to a (right) edge state around $E=0$ as shown in Fig.~\ref{Fig3}(c). The extent of the edge states is quantified by the factor $\eta^{}_{0}\equiv \sum^{5}_{i=1}|\mathbf{V}^{}_{i}|^{2}_{}-\sum^{N}_{i=N-4}|\mathbf{V}^{}_{i}|^{2}_{}$, with the left (right) edge state characterized by $\eta^{}_{ 0}\sim1$ ($ -1$).} On the other hand, the energy $\varepsilon^{}_{2}$ at which ${\cal F}^{}_{0,11}$ vanishes depends on the parameters  \cite{Supplementary}. In particular, as $\Phi$ increases, ``pushing" the system to flow from the ``AI'' class to the ``A" one passing through ``D",  that energy varies with $j^{}_{1}$, making the parameter range available for $\varepsilon^{}_{2}$  broader than that for the right endpoint.


For the  $\lambda=1$ configuration, the appearance of a right-side edge state necessitates
\begin{align}
{\cal F}^{}_{1,22}=& 0~~~\forall~~ \{E=\varepsilon^{}_{1},\ \ \varepsilon^{2}_{1} +j^{2}_{1} >j^{2}_{2} \}\ ,
 \end{align}
where $\varepsilon^{}_{1}\equiv\xi(j^{2}_{1}-j^{2}_{2})/(j^{2}_{1}+j^{2}_{3})$, i.e.,
the energy of the edge state is not the same in the various classes.
In particular, for the ``D'' or ``BDI'' class, the relevant parameter range is
$|j^{}_{2}/j^{}_{1}|<1$, opposite to that of the zero-energy edge state of the left side [i.e., $|j^{}_{2}/j^{}_{1}|>1$, see the second of Eqs.~(\ref{CY})]. It follows that there is always a zero-energy state in the band gap,   as seen in  Fig.~\ref{Fig3}(d).


In addition to the high-symmetric classes~\cite{Fulga2011},  scattering theory can also be used to effectively probe the edge states of a disordered chain belonging to a low-symmetry class. Figure~\ref{Fig4}(a) portrays the energy spectra of an isolated disordered chain coupled to leads in the two configurations of endpoints, $\lambda=0,1$.  As seen, with the change of the magnetic flux, the energy edge states can appear to be robust  (``protected") against a weak disorder as one changes the symmetry class, ``AI''$\rightarrow$ ``A''$\rightarrow$ ``D''$\rightarrow$...$\rightarrow$``AI'', and only a single one-sided edge state remained in the band gap of $\lambda=1$.  It is also interesting to note that the variation of the edge states of a disordered chain can be echoed by the change of the  $\pi$-phased reflections of the two sides, see  Fig.~\ref{Fig4}(b).

\vspace{3mm}
\noindent{\color{blue}\bf{Conclusions.---}} By studying the scattering matrix of a hybrid system comprising a finite triangle chain coupled with two semi-infinite leads, we could follow the conditions for the appearance of edge states and their relationships with the reflection amplitudes' phases in several symmetry classes, topological as well as non-topological. In particular, we show that a negative amplitude reflection can appear in non-topological classes, and can be correlated with the appearance of edge states, even in the presence of weak disorder, rendering these features as dubious indicators of a topological transition. Our study compares in detail the reflection amplitudes and edge states on both endpoints of a finite system, emphasizing the dependence on the length of the chain and the configuration of its coupling to the leads. Since scattering properties can be measured on finite chains (say, of cold atoms) at low temperatures,  and as we have found that (i) weak disorder is not detrimental for edge states and (ii) by applying Aharonov-Bohm flux the system can ``flow" in the space of symmetry classes, it is hoped that our study will invoke more interest regarding the characterization of topological materials via scattering properties.


\begin{acknowledgments}
This work is supported by the National Natural Science Foundation of China (Grant Nos. 92165208 and 11874071), and the Key-Area Research and Development Program of Guangdong Province (Grant No. 2020B0303060001).
\end{acknowledgments}





\end{document}


\title{Supplemental Material for \\``Scattering description of edge states  in  Aharonov-Bohm triangle chains"\\
}

\author{Zhi-Hai\!  Liu }
\email{liuzh@baqis.ac.cn}
\affiliation{Beijing Academy of Quantum Information Sciences, Beijing 100193, China }

\author{O. Entin-Wohlman}
\email{orawohlman@gmail.com}
\affiliation{School of Physics and Astronomy, Tel Aviv University, Tel Aviv 6997801, Israel}

\author{A. Aharony}
\affiliation{School of Physics and Astronomy, Tel Aviv University, Tel Aviv 6997801, Israel}

\author{J. Q. You}
\affiliation{School of Physics, Zhejiang University, Hangzhou 310027, China}

\author{H. Q. Xu}
\affiliation{Beijing Key Laboratory of Quantum Devices, Key Laboratory for the Physics and Chemistry of Nanodevices, and School of Electronics, Peking University, Beijing 100871, China}
\affiliation{Beijing Academy of Quantum Information Sciences, Beijing 100193, China }

\maketitle


 This ``supplemental material" presents two methodologies for studying topological properties of a triangle chain.
 The Bloch spectrum of a closed chain obeying periodic boundary conditions together with its topological features is analyzed in the first part. A finite chain is considered in the second part, exploiting the transfer-matrix approach. This transfer matrix allows for constructing the scattering matrix and analyzing its relation to the topological nature of the chain. In addition,  the transfer-matrix method is used to derive the scattering matrix of a disordered chain.

 \tableofcontents

\section{The Bloch Hamiltonian of a periodic triangle chain}
\label{SEC1}

 The triangle chain we consider is depicted in   Fig. 1(a)   of the main text.  It obeys the tight-binding Hamiltonian
\begin{align}
H^{}_{0}=-
\sum^{N-1}_{n=1}
\left(j^{}_{2}b^{\dagger}_{n}e^{i\phi^{}_{2}}_{}a^{}_{n+1}
+j^{}_{3}a^{\dagger}_{n}e^{i\phi^{}_{3}}_{}a^{}_{n+1}+{\rm h.c.}\right)-\sum^{N }_{n=1} \left(j^{}_{1}a^{\dagger}_{n}e^{i\phi^{}_{1}}_{}b^{}_{n}+{\rm h.c. }\right)\ ,
\label{H01}
\end{align}
where  $N$ is the number of unit cells~\cite{com}. In Eq. (\ref{H01}),   $a_{n}$ ($b_{n}$) is the annihilation operator of an electron residing on site $a$ ($b$) in the $n$th unit cell.  $j^{}_{1}$ denotes the intra-cell  tunneling amplitude,
$j^{}_{2}$ and $j^{}_{3}$ are the inter-cell amplitudes, all taken to be real.  The phases $\phi_{1}$, $\phi^{}_{2}$, and $\phi_{3}$ are the partial  Aharonov-Bohm phases accumulated on the tunneling amplitudes $j^{}_{1}$, $j^{}_{2}$, and $j^{}_{3}$, respectively, and  $\Phi=\phi^{}_{1}+\phi^{}_{2}-\phi^{}_{3}$ is the total magnetic flux penetrating each triangle (in units of the flux quantum).

In Fourier space,
the creation   (vector)  operator $\psi^{\dagger}_{n}=\{b^{\dagger}_{n}, a^{\dagger}_{n}\}$ becomes
$\psi^{\dagger}_{n}=N^{-1/2}_{}\sum^{}_{k'}\exp[ik^{\prime}_{}n]\psi^{\dagger}_{k^{\prime}_{}} $
with  $\psi^{\dagger}_{k^{\prime}_{}}=\{b^{\dagger}_{k^{\prime}_{}}, a^{\dagger}_{k^{\prime}_{}}\}$ and   $k^{\prime}_{}=2\pi \ell/N$, $\ell=0,1,\ldots ,N-1$.
The  Hamiltonian   is then
$ H^{}_{0}=\sum_{k'^{}_{}\in {\rm BZ}}\psi^{\dagger}_{k'_{}}  H^{}_{\rm B}(k')\psi^{}_{k'}\ ,
$
with
\begin{align}
H^{}_{\rm B}(k')= -\left[\begin{array}{cc}0&\ \
 j^{}_{1}e^{-i\phi^{}_{1}}+j^{}_{2}e^{i\phi^{}_{2}}e^{-ik'}\\
j^{}_{1}e^{i\phi^{}_{1}}+j^{}_{2}e^{-i\phi^{}_{2}}e^{ik'} &\ \  2j^{}_{3}\cos(k'-\phi^{}_{3})\end{array}
\right]\equiv\left [\begin{array}{cc}e^{-i\phi^{}_{1}/2}&0\\0&e^{i\phi^{}_{1}/2}\end{array}\right]
H(k)\left [\begin{array}{cc}e^{i\phi^{}_{1}/2}&0\\0&e^{-i\phi^{}_{1}/2}\end{array}\right]\ ,
 \end{align}
 and $k=\phi^{}_{1}+\phi^{}_{2}-k^{\prime}_{}$.
 The symmetry properties of the chain are encoded in the Hamiltonian
\begin{align}
H(k)= -\left[\begin{array}{cc}0 &\ \
 j^{}_{1}+j^{}_{2}e^{ik} \\
j^{}_{1}+j^{}_{2}e^{-ik} &\ \  2j^{}_{3}\cos(k-\Phi)\end{array}
\right]\equiv\mathbf{m}(k)  \cdot \boldsymbol{\sigma}-m^{}_{z}(k)\ ,
\label{HEK}
\end{align}
 which is gauge invariant and depends  only on the total flux $\Phi$. The $k-$dependent vector   $\mathbf{m} (k)=\{m^{}_{x}(k) , m^{}_{y} (k), m^{}_{z}(k) \}$  comprises the components  $m^{}_{x}(k)=-j^{}_{1}-j^{}_{2}\cos(k)$, $m^{}_{y}(k)=j^{}_{2}\sin(k)$, and $m^{}_{z}(k)=j^{}_{3}\cos(k-\Phi)$, and  $\boldsymbol{\sigma}=\{\sigma^{}_{x},\sigma^{}_{y},\sigma^{}_{z}\}$ is the vector of the  Pauli matrices. Here and below, scalar terms like  $m^{}_{z}(k)$ are  multiplied by the $2\times 2$ unit matrix.
 The  Bloch spectrum of the triangle chain  is
 \begin{align}
E^{}_{\pm}(k)=\pm m(k)-m^{}_{z}(k)\ ,\ \ \ m(k)=|\mathbf{m}(k)|=
[j^{2}_{1}+j^{2}_{2}+2j^{}_{1}j^{}_{2}\cos(k)+j^{2}_{3}\cos^{2}_{}(k-\Phi)]^{1/2}_{} \ .
\label{blochs}
 \end{align}

 As functions of its parameters, in particular, the next nearest-neighbor tunneling amplitude $j^{}_{3}$ and the magnetic flux $\Phi$, the Hamiltonian we consider belongs to different symmetry classes.  Recall that the chain hosts spinless electrons on a two-sublattice structure. In that case, the time-reversal operator ${\cal T}$  is simply $K$, the complex-conjugation operator, and consequently a time-reversal symmetric Bloch Hamiltonian obeys \cite{Chiu2016}
\begin{align}
{\cal T} H(k){\cal T}^{-1}= H(-k)\ .
\end{align}
The particle-hole operator ${\cal P}$ is $\sigma^{}_{z}K$, where $\sigma_{z}$ refers to the  ($a,b$) sites' space.  $H(k)$   is  particle-hole symmetric when
\begin{align}
{\cal P}H(k){\cal P}^{-1}=-H(-k)\ .
\end{align}
The chiral symmetry operator, ${\cal C}$, is the product   of the previous two symmetries,  i.e., ${\cal C}={\cal T}{\cal P}=\sigma^{}_{z}$,  such that
\begin{align}
{\cal C}H(k){\cal C}^{-1}=-H(k)\ .
\end{align}

 For   $j^{}_{3}=0$ the triangle chain is reduced to the    Su-Schrieffer-Heeger (SSH) model~\cite{Su1979}, which belongs to the  ``BDI"  symmetry  class~\cite{Chiu2016}, where all three symmetries are obeyed.  Chiral symmetry is broken when $j^{}_{3}\neq 0$ and the time-reversal symmetry is lifted for $\Phi\neq 0, \pm\pi$.  Thus, the triangle chain belongs to the ``AI" class when $\Phi=0$ and $j^{}_{3}\neq 0$.   Interestingly,    at $\Phi=\pm \pi/2$   particle-hole symmetry is restored for $j^{}_{3}\neq 0$, rendering the chain to be in the ``D"  class.
    For all other flux values and when $j^{}_{3}\neq 0$, the triangle chain belongs to the ``A" class, where all three symmetries are broken.


\begin{figure}
\centering
\includegraphics[width=0.760\textwidth]{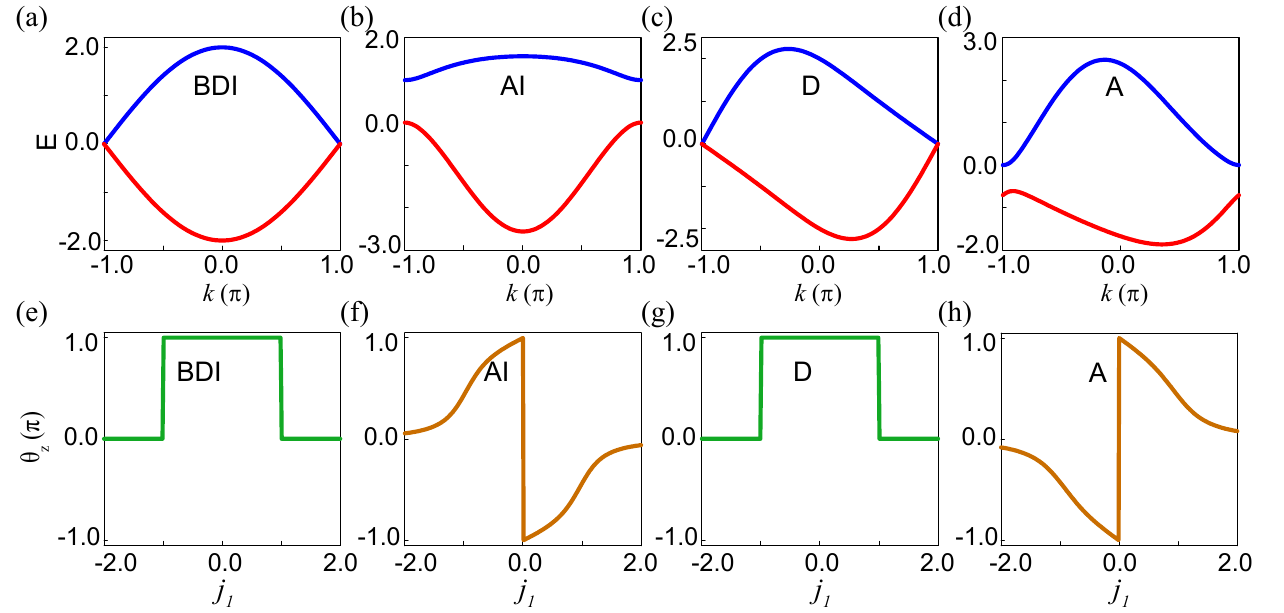}
\caption{(color online) (a)-(d) Energy spectra of the triangle chain as functions of  $k^{}_{}$   for  four symmetry  classes.  The  energy and tunneling amplitudes are in units  of $j^{}_{2}$, with $j^{}_{1}=1.0$ for all four classes.   In the ``BDI" class  $j^{}_{3}$, which breaks chiral symmetry, is zero.  It is chosen to be   $j^{}_{3}=0.5$  for the other three classes.  The magnetic flux $\Phi$   (in units of the flux quantum) vanishes for the ``BDI" and ``AI" classes (which are time-reversal symmetric), and is    $\Phi= 0.5\pi$, and  $\Phi=0.75\pi$  for the ``D" and ``A" classes, respectively. (e)-(h) The Zak phase  $\vartheta^{}_{\rm zak}$, defined  in Eq.~(\ref{ZAK}), as a function of $j^{}_{1}$ for the    corresponding symmetry  classes.  Other parameters are as in panels (a)-(d).  }
\label{Figs1}
\end{figure}



 For the triangle chain, the closing of the gap occurs for $E^{}_{+}(k)-E^{}_{-}(k)\equiv 2m^{}_{}(k)=0$, which  implies two conditions that have to be obeyed together,
 \begin{align}
(i)~~j^{}_{3}\cos^{}(k-\Phi)=0 ~~~~~{\rm and }~~~~~~(ii)~~j^{2}_{1}+j^{2}_{2}+2j^{}_{1}j^{}_{2}\cos (k)=0\ .
\label{conds}
\end{align}
 The second condition means that  $j^{}_{1}/j^{}_{2}=-\exp[\pm ik]$. Since the tunneling amplitudes are assumed to be real, it  is fulfilled for $j^{}_{1}=j^{}_{2}$ and $k=\pm\pi$, or for $j^{}_{1}=-j^{}_{2}$ with $k=0$. The first condition in Eq. (\ref{conds}) is obeyed for $j^{}_{3}=0$  (which reduces the triangle chain to  the particle-hole symmetric  SSH model), or for $k=\Phi+\pi/2$. In the first case, the chain belongs to the ``BDI" class, since it preserves time-reversal symmetry. In the second case (where  time-reversal symmetry is violated) the flux must assume the values $\pm\pi/2$ [in order to fulfill the first condition in Eq. (\ref{conds})],
 and the chain belongs to the ``D" class.
  For other values of $j^{}_{3}$ and $\Phi$, the gap in the Bloch spectrum does not close, as illustrated in Figs.~\ref{Figs1}(a)-\ref{Figs1}(d).


 The energy-band analysis of the triangle chain is compatible with the general statements of Ref.~\onlinecite{Ezawa2013}, showing that the bulk-boundary correspondence implies that closing of a gap is equivalent to a topological transition only if the transition happens within the same topological symmetry class, e.g., "BDI" and "D". This reference also indicated that this does not apply to the lower symmetry classes, e.g., "A" and "AI".

 Another hallmark of topological aspects  in the band structure is the Zak phase~\cite{Zak1989}, $\vartheta^{}_{\rm zak}$,
 \begin{align}
\vartheta^{}_{{\rm zak},\mp}=i\int^{\pi}_{-\pi} u^{\dagger}_{\mp}(k)\partial^{}_{k}u^{}_{\mp}(k) dk\ ,
\label{ZAK}
 \end{align}
 where $u^{}_{-/+}(k)$ are the Bloch functions  corresponding to
  the lower/upper energy band.  In topological phases, $\vartheta^{}_{\rm zak}/\pi$ is equal to the integer winding number of the vector ${\bf m}(k)$ around the Brillouin zone ($-\pi<k<\pi$)~\cite{Li2015}.     For instance, the Zak phase of the  SSH model in the ``BDI" class is quantized~\cite{Atala2013}.  To facilitate the calculation of the Zak    phase   in the general case, the vector $\mathbf{m}(k)$ is expressed in spherical coordinates,
 \begin{align}
&\mathbf{m}(k)=m(k)\{\cos( \varphi)\sin( \theta ), \sin( \varphi)\sin( \theta ),\cos( \theta )\}\ ,
\label{angles}
\end{align}
with  $\tan[\varphi(k)]=- j^{}_{2}\sin(k)/[j^{}_{1}+j^{}_{2}\cos(k)]$ and $ \cos[\theta(k)]=j^{}_{3}\cos(k-\Phi)/m(k)$ .
The Bloch Hamiltonian (\ref{HEK}) can now be rotated by  the unitary transformation
$\hat{\mathbf{R}}(\varphi,\theta)=\exp[i(\theta/2)\sigma^{}_{y}]\exp[i(\varphi/2)\sigma^{}_{z}]$,
\begin{align}
\hat{\mathbf{R}}^{}_{}H(k)\hat{\mathbf{R}}^{\dagger}_{}=m(k)\sigma_{z}-m^{}_{z}(k)\ ,
\end{align}
and  the Bloch functions of $H(k)$   are
\begin{align}
u^{}_{-}(k)=\hat{\mathbf{R}}^{\dagger}_{}\left[\begin{array}{c} 0\\1
\end{array}\right]=\left[\begin{array}{c}
\sin( \theta/2)e^{-i\varphi}_{}
\\
-\cos(\theta/2)\end{array}\right]\ ,  ~~~~~~~u^{}_{+}(k)=\hat{\mathbf{R}}^{\dagger}_{}\left[\begin{array}{c}1\\0\end{array}\right]=\left[\begin{array}{c}
\cos( \theta/2)e^{-i\varphi}_{}
\\
\sin(\theta/2)\end{array}\right]\ .
\end{align}
It follows that
\begin{align}
\vartheta^{}_{{\rm zak},\mp}=\int^{\pi}_{-\pi}dk \frac{1}{2}\left\{ 1\mp\cos[\theta(k)]\right\} \frac{\partial\varphi(k) }{\partial k}\ ,\ \ \ {\rm where}\ \ \ \frac{\partial\varphi}{\partial k}
=-\frac{j^{}_{2}[j^{}_{2}+j^{}_{1}\cos(k)]}{j^{2}_{1}+j^{2}_{2}+2j^{}_{1}j^{}_{2}\cos(k)}\ .
\label{zakx}
\end{align}
When $j^{}_{3}=0$  the vector ${\bf m}(k)$ lies in the $x-y$ plane, and the
 winding number is
\begin{align}
  \vartheta^{}_{\rm zak,\mp}/\pi=\begin{cases}
-1~~~~~~|j^{}_{2}|>|j^{}_{1}|\\
0~~~~~~~~~|j^{}_{2}|<|j^{}_{1}|\
\end{cases}\ ,
\label{cas}
\end{align}
as expected for the ``BDI" class.
For the ``D" class with $\Phi=\pm \pi/2$, it is found   that
 $\cos[\theta(k)]\partial^{}_{k}\varphi(k)$  is an odd function of $k$, and  the integral  $\int^{\pi}_{-\pi} \cos[\theta(k)]\partial^{}_{k} \varphi(k)dk$    is therefore equal to zero. Then  $\vartheta^{}_{\rm zak,\mp}$   takes the same quantized values as in the ``BDI" class, Eqs. (\ref{cas}).
For the ``AI" and  ``A" symmetry classes, characterized by a nonzero $j^{}_{3}$ and $\Phi\neq \pm \pi/2$, the integral of $\cos[\theta(k)]\partial^{}_{k}\varphi(k)$ over the Brillouin zone is nonzero, and the Zak phase assumes arbitrary values, as illustrated in Figs.~\ref{Figs1}(e)-\ref{Figs1}(h).     Indeed,  these two symmetry classes are considered `trivial', or non-topological, in which case Refs.~\onlinecite{Ezawa2013,Chiu2016} showed that the classes ``A'' and ``AI'' do not have a quantized invariant.


  \section{Scattering from a finite triangle chain}

In the main text, we consider the possibility of investigating topological features by using scattering theory~\cite{Fulga2012}.
To this end, two semi-infinite leads are attached to the two ends of a finite chain, as shown in Fig.~1(b)  of the main text. The discrete model of the hybrid system is detailed in the main text,   see Eq. (1) there and Ref. \onlinecite{com} below.   As discussed in the caption to  Fig. 1(b) of the main text, we consider two possible configurations of the coupling to the leads:
  $\lambda=0$   corresponds to the case where the finite chain comprises an integer number of unit cells and $\lambda=1$ is the configuration where the leads are attached to $a$ sites on both sides.  It turns out that this distinction leads to significantly different results.


\subsection{The transfer matrix  }

  We first give details of the derivation of   Eq. (4)  in the main text. The Schr\"{o}dinger equations at the inner sites of the chain are
\begin{align}
E\Psi^{}_{b}(n)=-j^{}_{1}e^{-i\phi^{}_{1}}\Psi^{}_{a}(n)-j^{}_{2}e^{i\phi^{}_{2}}\Psi^{}_{a}(n+1)\ ,\ \  1\leq n\leq N-1\ ,
\label{ESD}
\end{align}
and
\begin{align}
&E\Psi^{}_{a}(n+1)=-j^{}_{1}e^{i\phi^{}_{1}}\Psi^{}_{b}(n+1)-j^{}_{2}e^{-i\phi^{}_{2}}\Psi^{}_{b}( n )
-j^{}_{3}e^{i\phi^{}_{3}}\Psi^{}_{a}(n+2)-j^{}_{3}e^{-i\phi^{}_{3}}\Psi^{}_{a}(n)\ ,\ \ \  1\leq n \leq N-2\ ,
\label{ESD2}
\end{align}
where $\Psi^{}_{a}(n)$ and $\Psi^{}_{b}(n)$   are the wave functions of the electronic states at energy $E=-2J^{}_{0}\cos(\kappa)$ at sites  $a$ and $b$  of  the  $n$th cell.
The Schr\"{o}dinger equations at the two endpoints of the chain are determined by the boundary conditions, as discussed below.
 Rewriting Eqs.~(\ref{ESD}) and (\ref{ESD2}) as
\begin{align}
\Psi^{}_{a}(n+1)=-\frac{e^{-i\phi^{}_{2}}}{j^{}_{2}}\Big(E\Psi^{}_{b}(n)+j^{}_{1}e^{-i\phi^{}_{1}}\Psi^{}_{a}(n)\Big )\ ,
\label{psia}
\end{align}
and
\begin{align}
\Psi^{}_{b}(n+1)=-\frac{e^{-i\phi^{}_{1}}}{j^{}_{1}}\Big (E\Psi^{}_{a}(n+1)+j^{}_{2}e^{-i\phi^{}_{2}}\Psi^{}_{b}(n)+j^{}_{3}e^{i\phi^{}_{3}}\Psi^{}_{a}(n+2)+j^{}_{3}e^{-i\phi^{}_{3}}\Psi^{}_{a}(n)\Big )\ ,
\label{psib}
\end{align}
  and using  Eq. (\ref{psia})  for $\Psi^{}_{a}(n+2)$,  Eq.~(\ref{psib}) becomes
\begin{align}
W\Psi^{}_{b}(n+1)=-e^{-i\phi^{}_{1}}\Big (  [Ej^{}_{2}-j^{}_{1}j^{}_{3}e^{-i\Phi} ]\Psi^{}_{a}(n+1)+j^{2}_{2}e_{}^{-i\phi_{2}}\Psi^{}_{b}(n)+j^{}_{3}j^{}_{2}e^{-i\phi_{3}}\Psi_{a}(n)\Big )\ ,
\label{psib2}
\end{align}
where
\begin{align}
W^{ }= j^{}_{1}j^{}_{2}-Ej^{}_{3}e^{-i\Phi}\equiv e^{i\beta}|W|\ .
\label{W}
\end{align}
Then,   substituting Eq.~(\ref{psia})   in  Eq.~(\ref{psib2}),  one   finds
\begin{align}
We^{ i(\Phi+\phi^{}_{3})}\Psi^{}_{b}(n+1)&=\Big(E^{2}-j^{2}_{1}-j^{2}_{2}+\frac{j^{}_{1}}{j^{}_{2}}W\Big)\Psi^{}_{b}(n)+e^{-i \phi^{}_{1} }
\Big(-j^{}_{3}j^{}_{2}e^{i\Phi}+Ej^{}_{1}
-\frac{j^{2}_{1}j^{}_{3}}{j^{}_{2}}e^{-i\Phi}\Big)\Psi^{}_{a}(n)\ .
\label{secf}
\end{align}
  Combining  Eqs.~(\ref{psia}) and (\ref{secf})  leads to the transfer matrix equation
  \begin{align}
  \mathbf{V}^{}_{n+1}={\cal T}(E)\mathbf{V}^{}_{n}\ ,
  \end{align}
  with $\mathbf{V}^{}_{n}=\{\Psi^{}_{b}(n),\Psi^{}_{a}(n)\}^{T}_{}$ and ${\cal T}(E)$ given in Eq. (4) of the main text
  \begin{align}
  {\cal T}(E)= \exp(\alpha  \hat{\mathbf{v}}\cdot \boldsymbol{\sigma}^{}_{}-i\zeta)\ .\label{TEM}
  \end{align}
Here, $\boldsymbol{\sigma}=(\sigma^{}_{x},\sigma^{}_{y},\sigma^{}_{z})$ is the vector of the Pauli matrices,   $\zeta=\Phi+\phi^{}_{3}+\beta$, $
 \cosh(\alpha)= (E^{2}_{}-j^{2}_{1}-j^{2}_{2})/(2|W|)\ ,
$
  and
  \begin{align}
  \hat{\bf v}=\frac{e^{i(\phi^{}_{1}+\beta)}_{}}{2j^{}_{2}\sinh(\alpha)}\left\{w^{}_{1}-E,i(w^{}_{1}+E),
w^{}_{2}+e^{-i\phi^{}_{1}}_{}j^{}_{1}\right\}\ ,
  \end{align}
with
$w^{}_{1}=e^{-2i\phi^{}_{1}}_{}[Ej^{}_{1}j^{}_{2}-j^{}_{3}(j^{2}_{1}e^{-i\Phi}_{}+j^{2}_{2}e^{i\Phi}_{})]/W$, and
$w^{}_{2}=e^{-i\phi^{}_{1}}_{}(2j^{}_{2}e^{-i\beta}_{}\cosh\alpha+j^{}_{1})$.


\subsubsection{The periodic chain }

\begin{figure}
\centering
\includegraphics[width=0.660\textwidth]{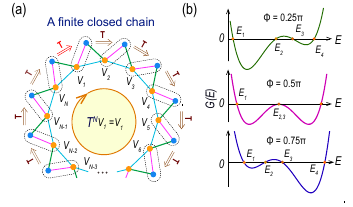}
\caption{(color online) (a) Schematic diagram of the transfer relation between two nearest-neighboring sites in a periodic triangle chain. The periodic boundary condition for a chain of $N$ sites leads to the relation ${\cal T}^{N}_{} \mathbf{V}^{}_{1}=\mathbf{V}^{}_{1} $. (b) The  function  $G(E)\equiv(E^{2}_{}-j^{2}_{1}-j^{2}_{2})^{2}_{}-4|W|^{2}_{}$ which specifies   the periodic-chain spectrum, as a function of $E$  for   $j^{}_{1}=j^{}_{2}=1.0$, $j^{}_{3}=0.5$, and different  values of the Aharonov-Bohm flux  $\Phi$ (energies are in units of $j^{}_{2}$).  The  four   energies where $G(E)=0$ are   $E^{}_{i=1-4}$ with $E^{}_{1}< E^{}_{2}\leq E^{}_{3}<E^{}_{4}$;   the spectrum is gapped for $E_{3}<E<E^{}_{2}$.     }
\label{Figs2}
\end{figure}

The Bloch spectrum,  Eq.~(\ref{blochs}),  can be derived from the transfer matrix relation applied for a  closed chain obeying periodic boundary condition, as shown in Fig.~\ref{Figs2}(a).   In that case, the Hamiltonian is
\begin{align}
H^{}_{\rm p}=& H^{}_{0}
 + \left(j^{}_{2}b^{\dagger}_{N}e^{i\phi^{}_{2}}_{}a^{}_{1}
+j^{}_{3}a^{\dagger}_{N}e^{i\phi^{}_{3}}_{}a^{}_{ 1}+{\rm h.c.} \right)\ ,
\end{align}
where $H^{}_{0}$ is given in Eq.~(\ref{H01}).
  The relations among the  vectors   at the two endpoints, $\mathbf{V}^{}_{1}$ and $\mathbf{V}^{}_{N}$,   and the two nearest-neighboring vectors $\mathbf{V}^{}_{N-1}$ and $\mathbf{V}_{N}$, are found from the    Schr\"{o}dinger equations at the $a$ and  $b$-sites of $N-$th unit cell
\begin{align}
&\Psi^{}_{a}(1)=-\frac{e^{-i\phi^{}_{2}}}{j^{}_{2}}\Big (E\Psi^{}_{b}(N)+j^{}_{1}e^{-i\phi^{}_{1}}
\Psi^{}_{a}(N)\Big )\ ,\nonumber\\
 &E\Psi^{}_{a}(N)=-j^{}_{1}e^{i\phi^{}_{1}}\Psi^{}_{b}(N)-j^{}_{3}e^{i\phi^{}_{3}}\Psi^{}_{a}(1)-j^{}_{3}e^{-i\phi^{}_{3}}\Psi^{}_{a}(N-1)
 -j^{}_{2}e^{-i\phi^{}_{2}}\Psi^{}_{b}(N-1)\ .
 \label{ESD4}
\end{align}
 Eliminating $\Psi^{}_{a}(1)$ from the second of Eqs.~(\ref{ESD4}) by using the first and then removing $\Psi^{}_{a}(N)$ by using the Schr\"{o}dinger equation for $\Psi^{}_{b}(N-1)$, i.e.,
 \begin{align}
 E\Psi^{}_{b}(N-1)=-j^{}_{1}e^{-i\phi^{}_{1}}\Psi^{}_{a}(N-1)-j^{}_{2}e^{i\phi^{}_{2}}\Psi^{}_{a}(N)\ , \label{BN-1}
 \end{align}
   yields
$
 \Psi^{}_{b}(N)={\cal T}^{}_{11} \Psi^{}_{b}(N-1)+{\cal T}^{}_{12} \Psi^{}_{a}(N-1)
$.
 Together with Eq.~(\ref{BN-1}), one  then finds $\mathbf{V}^{}_{N}={\cal T}^{}_{}\mathbf{V}_{N-1}$.    At the other  endpoint  of the chain, the Schr\"{o}dinger  equation at the  $a$-site of the first cell reads
\begin{align}
 &E\Psi^{}_{a}(1)=-j^{}_{1}e^{i\phi^{}_{1}}\Psi^{}_{b}(1)-j^{}_{3}e^{i\phi^{}_{3}}\Psi^{}_{a}(2)-j^{}_{3}e^{-i\phi^{}_{3}}\Psi^{}_{a}(N)
 -j^{}_{2}e^{-i\phi^{}_{2}}\Psi^{}_{b}(N)
 \label{ESD3}\ .
 \end{align}
 Using  Eq.~(\ref{ESD}) for $\Psi^{}_{b}(1)$ to eliminate $\Psi^{}_{a}(2)$, i.e.,
 \begin{align}
  &  e^{i\phi^{}_{1}}  W\Psi^{}_{b}(1)=  (j^{}_{3} j^{}_{1} e^{ -i\Phi}-Ej^{}_{2})\Psi^{}_{a}(1)-j^{}_{2}j^{}_{3}e^{-i\phi^{}_{3}}\Psi^{}_{a}(N )
 -j^{2}_{2}e^{-i\phi^{}_{2}}\Psi^{}_{b}(N)
 \end{align}
  and combined with  the first of Eqs.~(\ref{ESD4}),   we  obtain $\mathbf{V}^{}_{1}={\cal T}\mathbf{V}^{}_{N}$.  On the other hand
  \begin{align}
\mathbf{V}^{}_{1}={\cal T}\mathbf{V}^{}_{N}={\cal T}[{\cal T}\mathbf{V}^{}_{N-1}]={\cal T}^{N}_{}\mathbf{V}^{}_{1}\ ,
\label{per}
\end{align}
 leading to  $\left({\cal T}^{N}_{}-1\right)\mathbf{V}^{}_{1}=1$.  Exploiting Eq. (3) in the main text,  the matrix ${\cal T}^{N}-1$ takes the form
 \begin{align}
{\cal T}^{N}_{}-1=&e^{-iN\zeta} \left\{ \big[\cosh(N\alpha)
+\sinh(N\alpha)\hat{\bf  v}\cdot\sig\big]-e^{iN\zeta}\right\}\ ,
\end{align}
 whose  determinant  is
$\det[{\cal T}^{N}_{}-1]= 2e^{-iN\zeta}_{} [\cos(N\zeta)-\cosh(N\alpha)]$.
It follows that the spectrum of the periodic chain is given by
$\cos(N\zeta)=\cosh(N\alpha)$, as discussed in the main text.
 Here, the analytical expression  for $\exp[\pm\alpha ]$   is
\begin{align}
\exp\left[\pm\alpha\right]=\frac{1}{2|W|} \left[E^{2}_{}-j^{2}_{1}-j^{2}_{2}\mp\sqrt{G(E)}\right]\ ,
\label{exp}
 \end{align}
 with $G(E)=(E^{2}_{}-j^{2}_{1}-j^{2}_{2})^{2}_{}- 4|W|^{2}_{}$.

For a bulk state of a  finite chain,   the exponential factor  $\alpha$  must be purely imaginary
since $|\cosh(N\alpha)|=|\cos(N\zeta) |\leq 1$. This means that
 $G(E)<0$ based on Eq.~(\ref{exp}). The borderline of the bulk energy bands is then determined by solving the equation  $G(E)=0$, i.e., $(E^{2}_{}-j^{2}_{1}-j^{2}_{2})^{2}_{}= 4|W|^{2}_{}$,  and the band gap is the difference between the two middle energy-solutions of the quartic formula [see Fig.~\ref{Figs2}(b)].
Varying the flux $\Phi$ threading the triangle shows that the band gap cannot close in the ``A'' or ``AI'' classes, see Fig.~\ref{Figs2}(b).  The band gap does close in the ``D'' class with  $|j^{}_{1}/j^{}_{2}|=1$ and also in the  ``BDI'' class ($j^{}_{3}=0$).   This is consistent with the results obtained from the Bloch band-theory in Figs.~\ref{Figs1}(a)-\ref{Figs1}(d).

  When $E$ is located  in the band gap,  $G(E)>0$ and $E^{2}_{}<j^{2}_{1}+j^{2}_{2}$. Equation (\ref{exp}) then implies that $\exp[\alpha]$ is real, and $\exp[\alpha]< -1$, so that $\alpha$ must have the form  $\alpha=i\pi+\beta^{\prime}_{}$, with $\beta^{\prime}_{}>0$.


 \subsubsection{The open chain}

 For the open chain, the amplitudes of the first and the $(N-1)$th  cells  are still related by
 \begin{align}
 \mathbf{V}^{}_{N-1}={\cal T}^{N-2}_{} \mathbf{V}^{}_{1}\ ,
 \label{Vf1}
 \end{align}
  where ${\cal T}$ is the transfer matrix.
The transfer relation between the amplitudes of the two rightmost sites is different and depends  on which site is connected to the right-side lead.

  When the rightmost site is a $b$ site (and the chain consists of an integer number of unit cells, a configuration denoted in the main text by $\lambda=0$,  then   using Eq.~(\ref{ESD}) for $\Psi^{}_{b}(N-1)$, one   finds
   \begin{align}
   \Psi^{}_{a}(N)= -\frac{E}{j^{}_{2}}e^{-i\phi^{}_{2}}_{}\Psi^{}_{b}(N-1)-\frac{j^{}_{1}}{j^{}_{2}}e^{-i(\phi^{}_{1}+\phi^{}_{2})}_{}\Psi^{}_{a}(N-1)\ .  \label{Ea1}
   \end{align}
  However, contrary  to Eq.~(\ref{ESD2}), the Schr\"{o}dinger    equation   at the $a$  site  of   the  $N$th cell is
 \begin{align}
 E\Psi^{}_{a}(N)=&-j^{}_{1}e^{i\phi^{}_{1}}\Psi^{}_{b}(N)-j^{}_{2}e^{-i\phi^{}_{2}}\Psi^{}_{b}(N-1)
-j^{}_{3}e^{-i\phi^{}_{3}}\Psi^{}_{a}(N-1)\ . \label{an}
 \end{align}
 Using Eq.~(\ref{Ea1}) to eliminate $\Psi^{}_{a} (N)$, this becomes
 \begin{align}
  \Psi^{ }_{b} (N)= \frac{E^{2}_{}-j^{2}_{2}}{j^{}_{1}j^{}_{2}} e^{-i(\phi^{}_{1}+\phi^{}_{2})}_{}\Psi^{}_{b}(N-1)+\frac{Ej^{}_{1}-j^{}_{2}j^{}_{3}e^{i\Phi}_{}}{j^{}_{1}j^{}_{2}}e^{-i(2\phi^{}_{1}+\phi^{}_{2})}_{}
  \Psi^{}_{a}(N-1)\ .
 \end{align}
 Combining with Eq.~(\ref{an}), the transfer matrix relation $\mathbf{V}^{}_{N}={\cal M}^{}_{0}\mathbf{V}^{}_{N-1}$ is  obtained, with
 \begin{align}
 {\cal M}^{}_{0}=
 \frac{e^{-i(\phi^{}_{1}+\phi^{}_{2})}_{}}{j^{}_{1}j^{}_{2}}\left[\begin{array}{cc}
 E^{2}_{}-j^{2}_{2} &e^{-i\phi^{}_{1}}_{}(Ej^{}_{1}-j^{}_{2}j^{}_{3}e^{i\Phi}_{})\\
 -Ej^{}_{1}e^{i\phi^{}_{1}}_{}&-j^{2}_{1}
 \end{array}\right]\ .
 \end{align}
 The relation  between the amplitudes at the two endpoints  of the chain is then
\begin{align}
\left[\begin{array}{c}
\Psi^{}_{b}(N)\\
\Psi^{}_{a}(N)
\end{array}\right]={\cal X}^{}_{0}\left[\begin{array}{c}
\Psi^{}_{b}(1)\\
\Psi^{}_{a}(1)
\end{array}\right]\ ,\label{P0}
\end{align}
with
\begin{align}
{\cal X}^{}_{0}\equiv{\cal M}^{}_{0}{\cal T}^{N-2}_{}=\frac{e^{-i(N-1)\zeta}_{}e^{i\beta}_{}}{j^{}_{1}j^{}_{2}S^{}_{1}}\left[\begin{array}{cc}
  |W|S^{}_{N}+ j^{2}_{1}S^{}_{N-1}& \frac{e^{-i\phi^{}_{1}}_{}}{E}(j^{}_{1}[|W|S^{}_{N}+j^{2}_{1}S^{}_{N-1}]+j^{}_{2}e^{-i\beta}_{}[|W|S^{}_{N-1}+j^{2}_{1}S^{}_{N-2}])\\ \\
 -e^{i\phi^{}_{1}}_{}E^{ }_{}j^{}_{1}S^{}_{N-1}&-j^{2}_{1} S^{}_{N-1}-e^{-i\beta}_{}j^{}_{1}j^{}_{2}S^{}_{N-2}
\end{array}\right]\  ,  \label{X0}
\end{align}
and $S^{}_{n}=\sinh(n\alpha)$.


 In the  $\lambda=1$ configuration, which ends with an $a$ site, the $N$th cell contains a single site, i.e., the $a$-site,  and the relation between the amplitudes of the two rightmost sites is given by Eq.~(\ref{Ea1}).     Furthermore,  based on Eq.~(\ref{Vf1}), and because the rightmost cell  contains a single site, the transfer matrix between the endpoints of the chain is
\begin{align}
\left[\begin{array}{c}
\Psi^{}_{a}(N)\\
\Psi^{}_{b}(N-1)
 \end{array}\right]={\cal X}^{}_{1}\left[\begin{array}{c}
\Psi^{}_{b}(1)\\
\Psi^{}_{a}(1)
 \end{array}\right]\ ,\label{P1}
\end{align}
with
\begin{align}
{\cal X}^{}_{1}=\frac{e^{-i(N-1)\zeta}_{}e^{i\beta}_{}}{j^{}_{1}j^{}_{2}S^{}_{1}}\left[\begin{array}{cc}
 -e^{i\phi^{}_{1}}_{}E^{ }_{}j^{}_{1}S^{}_{N-1}&-j^{2}_{1} S^{}_{N-1}-e^{-i\beta}_{}j^{}_{1}j^{}_{2}S^{}_{N-2}\\ \\
e^{i(\phi^{}_{1}+\phi_{2})}_{} (j^{}_{1}j^{}_{2}S_{N-1}+e^{i\beta}_{}j^{2}_{1}S^{}_{N-2}) &~~~\frac{e^{i\phi^{}_{2}}_{}}{|W|}(-j^{}_{3}j^{}_{2}e^{i\Phi}+Ej^{}_{1}
-\frac{j^{2}_{1}j^{}_{3}}{j^{}_{2}}e^{-i\Phi})j^{}_{1}j^{}_{2}S^{}_{N-2}
\end{array}\right]\ . \label{X1}
\end{align}

\begin{figure}
\centering
\includegraphics[width=0.760\textwidth]{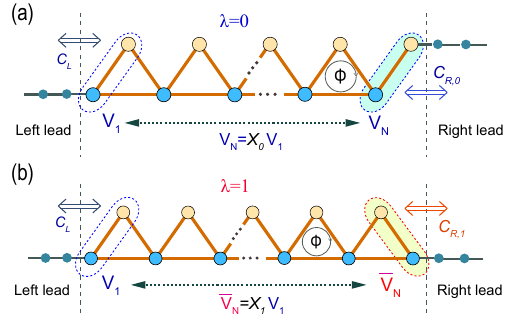}
\caption{(color online)   Schematic diagram of a finite chain coupled to two semi-infinite leads. (a)  represents the case when the finite chain is in the $\lambda=0$ configuration, in which the left (right) lead is coupled to a $a$ ($b$) sublattice of the chain.  Here, ${\cal X}^{}_{0}$ denotes the matrix interconnecting the amplitudes of the endpoints of the chain, $\mathbf{V}^{}_{1}$ and $\mathbf{V}^{}_{N}$ , see Eq.~(\ref{P0}). $C^{}_{L}$  and  $C^{}_{R,0}$ are the matrices responsible for the connection to the leads. (b) The $\lambda=1$ configuration,  in which the leads of both sides are coupled to   $a$ sublattices,   the matrix interconnecting the amplitudes of the endpoints is given by  ${\cal X}^{}_{1}$ in Eq.~(\ref{X1}). Due to the lack of a $b$-sublattice at the rightmost site,  the rightmost amplitude vector is $\overline{\mathbf{V}}^{}_{N}= \{\Psi^{}_{a}(N),\Psi^{}_{b}(N-1)\}$,   with   $C^{}_{R,1}$ being the relevant matrix for coupling to the right lead.    }
\label{Figs3}
\end{figure}


\subsection{Scattering matrix of a finite chain }


Once coupled to semi-infinite leads, the electronic states of the chain are mixed with the wave functions on the leads.    The  amplitude  at the endpoint of the left lead is $A^{}_{L}+B^{}_{L}$, where $A^{}_{L}$ and $B^{}_{L}$ are the amplitudes of the incoming and outgoing waves, obeying
\begin{align}
E\Psi^{}_{a}(1)=-J^{}_{\rm c}(A^{}_{L}+B^{}_{L})-j^{}_{1}e^{i\phi^{}_{1}}_{}\Psi^{}_{b}(1)-j^{}_{3}e^{i\phi^{}_{3}}_{}\Psi^{}_{a}(2)\ .
\end{align}
Eliminating $\Psi^{}_{a}(2)$ by using Eq.~(\ref{psia}) for $n=1$, one finds
\begin{align}
\Psi^{}_{b}(1)=\frac{j^{}_{2}e^{-i\phi^{}_{1}}_{}}{W}\left[-J^{}_{\rm c}(A^{}_{L}+B^{}_{L})-(E-\frac{j^{}_{1}j^{}_{3}}{j^{}_{2}}e^{-i\Phi}_{})\Psi^{}_{a}(1)\right]\ .
\end{align}
 Consequently,    the amplitude  vector $\mathbf{V}^{}_{1}$ becomes
\begin{align}
\left[\begin{array}{c}
\Psi^{}_{b}(1)\\
\Psi^{}_{a}(1)
 \end{array}\right]=  C ^{}_{L}\left[\begin{array}{c}
\Psi^{}_{a}(1)\\
J^{}_{\rm c} (A^{}_{L}+B^{}_{L})
 \end{array}\right]\ , ~~~{\rm with } ~~~  C ^{}_{L}=\frac{1}{W}\left[\begin{array}{cc}
 (j^{}_{1}j^{}_{3}e^{-i\Phi}_{}-Ej^{}_{2})e^{-i\phi^{}_{1}}_{}&- j^{}_{2}e^{-i\phi^{}_{1}}_{} \\
 W&0
 \end{array}\right]\ .\label{CL0}
\end{align}
 On the other side of the triangle chain the relation between the amplitudes of the incoming and outgoing waves, $A_{R}$ and $B_{R}$, and the states on the chain, depend on the configuration, see Fig.~\ref{Figs3}.   For the $\lambda=0$ case,  the amplitudes of the incoming and outgoing waves of the  right lead, $A^{}_{R}$ and $B^{}_{R}$, are  related  to $\Psi^{}_{b}(N)$ by
\begin{align}
\Psi^{}_{a}(N)=\frac{e^{i\phi^{}_{1}}_{}}{j^{}_{1}}\left[-J^{}_{\rm c}(A^{}_{R}+B^{}_{R})-E\Psi^{}_{b}(N)\right]\ .
\end{align}
The  vector on the left side of Eq.~(\ref{P0}) changes to
\begin{align}
\left[\begin{array}{c}
\Psi^{}_{b}(N)\\
\Psi^{}_{a}(N)
\end{array}\right]=  C ^{}_{R,0}\left[\begin{array}{c}
J^{}_{\rm c}(A^{}_{R}+B^{}_{R})\\
\Psi^{}_{b}(N)
\end{array}\right]\ ,~~~~{\rm with }~~~  C ^{}_{R,0}=\frac{1}{j^{}_{1}}\left[\begin{array}{cc}
0& j^{}_{1}
 \\
 - e^{i\phi^{}_{1}}_{}&-E e^{i\phi^{}_{1}}_{}
 \end{array}\right]\ .
 \label{CR0}
\end{align}
  For the $\lambda=1$ case,  since
\begin{align}
E\Psi^{}_{a}(N)=-j^{}_{2}e^{-i\phi^{}_{2}}_{}\Psi^{}_{b}(N-1)-J^{}_{\rm c}(A^{}_{R}+B^{}_{R})-j^{}_{3}e^{-i\phi^{}_{3}}_{}\Psi^{}_{a}(N-1)\ ,
\end{align}
  one finds  that
\begin{align}
\Psi^{}_{b}(N-1)=\frac{e^{i\phi^{}_{2}}_{}}{W^{\ast}_{}}\left[-j^{}_{1}J^{}_{\rm c}(A^{}_{R}+B^{}_{R})+(j^{}_{2}j^{}_{3}e^{i\Phi}_{}-j^{}_{1}E)\Psi^{}_{a}(N)\right]\ ,
\end{align}
in which   $\Psi^{}_{a}(N-1)$ is eliminated by using Eq.~(\ref{ESD}) for $\Psi^{}_{b}(N-1)$.
As a result, the wave vector on the left side of Eq.~(\ref{P1}) becomes
\begin{align}
\left[\begin{array}{c}
\Psi^{}_{a}(N)\\
\Psi^{}_{b}(N-1)
 \end{array}\right]=  C ^{}_{R,1}\left[\begin{array}{c}
J^{}_{\rm c}(A^{}_{R}+B^{}_{R})\\
\Psi^{}_{a}(N)
\end{array}\right]~~~{\rm with}~~~  C ^{}_{R,1}=\frac{e^{i\phi^{}_{2}}_{}}{W^{\ast}_{}}\left[\begin{array}{cc}
0&~~~~~~~~W^{\ast}_{}e^{-i\phi^{}_{2}}_{}\\
-j^{}_{1}&~~~~j^{}_{2}j^{}_{3}e^{i\Phi}_{}-j^{}_{1}E
\end{array}\right]\ . \label{CR1}
\end{align}
  Hence, using
  Eqs.~(\ref{CL0}), (\ref{CR0}), and (\ref{CR1}), the  transfer relations in Eqs.~(\ref{P0}) and (\ref{P1})  become
\begin{align}
  C^{}_{R,\lambda} \left[\begin{array}{c}
J^{}_{\rm c}(A^{}_{R}+B^{}_{R})\\
\Psi^{}_{\chi}(N)
\end{array}\right]={Z}^{}_{\lambda} \left[\begin{array}{c}
\Psi^{}_{a}(1)\\
J^{}_{\rm c}(A^{}_{L}+B^{}_{L})
\end{array}\right]\ ,
\end{align}
with ${  Z}^{}_{\lambda}={\cal X}^{}_{\lambda}   C ^{}_{L}$. aaa{question: should $C^{}_L$ not be $C^{}_{L,\lambda}$?} Then, it follows that
\begin{align}
{\cal F}^{}_{\lambda}\left[\begin{array}{c}
\Psi^{}_{a}(1)\\
\Psi^{}_{\chi}(N)
\end{array}\right]=J^{}_{\rm c}\left[\begin{array}{c}
A^{}_{L}+B^{}_{L}\\
 A^{}_{R}+B^{}_{R}
\end{array}\right]\ , \label{FJ}
\end{align}
with the four  matrix elements of ${\cal F}^{}_{\lambda}$  given by
\begin{align}
{\cal F}^{}_{\lambda,11}=&\frac{ [C^{}_{R,\lambda} ]^{}_{21}[Z^{}_{\lambda}]_{11}-[C^{}_{R,\lambda}]^{}_{11}[Z^{}_{\lambda}]_{21} }{[C^{}_{R,\lambda}]^{}_{11}[Z^{}_{\lambda}]_{22}-[C^{}_{R,\lambda}]^{}_{21}[Z^{}_{\lambda}]_{12} } ~~~~~~~ {\cal F}^{}_{\lambda,12}=\frac{ \det[  C^{}_{R,\lambda}] }{[C^{}_{R,\lambda}]^{}_{11}[Z^{}_{\lambda}]_{22}-[C^{}_{R,\lambda}]^{}_{21}[Z^{}_{\lambda}]_{12} }\nonumber\\
 {\cal F}^{}_{\lambda,21}=&\frac{ \det[Z^{}_{\lambda}]}{[C^{}_{R,\lambda}]^{}_{11}[Z^{}_{\lambda}]_{22}-[C^{}_{R,\lambda}]^{}_{21}[Z^{}_{\lambda}]_{12} } ~~~~~~~ {\cal F}^{}_{\lambda,22}= \frac{ [C^{}_{R,\lambda} ]^{}_{22}[Z^{}_{\lambda}]_{12}-[C^{}_{R,\lambda}]^{}_{12}[Z^{}_{\lambda}]_{22} }{[C^{}_{R,\lambda}]^{}_{11}[Z^{}_{\lambda}]_{22}-[C^{}_{R,\lambda}]^{}_{21}[Z^{}_{\lambda}]_{12} }\ .
\end{align}
  Exploiting the expressions for   $C^{}_{L}$, $C^{}_{R,\lambda}$, and ${\cal X}^{}_{\lambda}$ [see Eqs.~(\ref{X0}),~(\ref{X1}),~(\ref{CL0}), (\ref{CR0}), and (\ref{CR1})],   the matrices belonging to the  two ending configurations  are
  \begin{align}
{\cal F}^{}_{0}=&- \frac{1}{ |W|S^{}_{N}+j^{2}_{1}S^{}_{N-1}}\left[\begin{array}{cc}
 E^{ }_{}|W|S^{}_{N}+\frac{j^{2}_{1}j^{2}_{2}-|W|^{2}_{}}{E}S^{}_{N-1}  & \ \ \ j^{}_{1}S^{}_{1}|W|e^{i(\varphi^{}_{\rm t}+\phi^{}_{1})}_{} \\ \\
 j^{}_{1}S^{}_{1}|W|e^{-i(\varphi^{}_{\rm t}+\phi^{}_{1})}_{}   & E|W|S^{}_{N}
\end{array}\right]\ ,\nonumber\\
{\cal F}^{}_{1}=&- \frac{1}{ES^{}_{N-1}}\left[\begin{array}{cc}
 |W|S^{}_{N}+j^{2}_{2}S^{}_{N-1}  &-  |W|S^{}_{1}e^{i\varphi^{}_{\rm t}}_{}  \\ \\
-  |W|S^{ }_{1}e^{-i\varphi^{}_{\rm t}}_{} & |W|S^{}_{N}+j^{2}_{1}S^{}_{N-1}
\end{array}\right]\ , \label{FMS}
\end{align}
with  $\varphi^{}_{\rm t}=(N-1)\zeta$.
Note that for   the isolated chain, i.e., for  $J^{}_{\rm c}=0$, Eq.~(\ref{FJ})  gives
 \begin{align}
 {\cal F}^{}_{\lambda}\left[\begin{array}{c}
\Psi^{}_{a}(1)\\
\Psi^{}_{\chi}(N)
\end{array}\right]= 0\ ,
 \end{align}
  yielding for    the energy spectrum of the   isolated  chain
\begin{align}
\det[{\cal F}^{}_{\lambda}]=0\ . \label{det0}
\end{align}
  The  determinants  of ${\cal F}^{}_{0}$  and ${\cal F}^{}_{1}$   are
\begin{align}
\det[{\cal F}^{}_{0}]= |W|\frac{|W|S^{}_{N+1}+j^{2}_{2}S^{}_{N}}{|W|S^{}_{N}+j^{2}_{1}S^{}_{N-1}}\ ~~~{\rm and }~~~\det[{\cal F}^{}_{1}]= \frac{ (j^{2}_{1}j^{2}_{2}-|W|^{2}_{})S^{}_{N-1}+E^{2}_{}|W|S^{}_{N}}{E^{2}_{}S^{}_{N-1}}\ . \label{detF}
\end{align}

 For  nonzero $J^{}_{\rm c}$,
\begin{align}
E (A^{}_{L}+B^{}_{L})=-J^{}_{0}(A^{}_{L}e^{-i\kappa}_{}+B^{}_{L}e^{i\kappa}_{})-J^{}_{\rm c}\Psi^{}_{a}(1)~~~{\rm and}~~~ E (A^{}_{R}+B^{}_{R})=-J^{}_{0}(A^{}_{R}e^{-i\kappa}_{}+B^{}_{R}e^{i\kappa}_{})-J^{}_{\rm c}\Psi^{}_{\chi}(N)\ ,
\end{align}
 and one   obtains  that
\begin{align}
\left[\begin{array}{c}
\Psi^{}_{a}(1)\\
\Psi^{}_{\chi}(N)
\end{array}\right]=2i\frac{J^{}_{0}}{J^{}_{\rm c }}\sin(\kappa)\left[\begin{array}{c}
A^{}_{L} \\
A^{}_{R}
\end{array}\right]+\frac{J^{}_{0}}{J^{}_{\rm c}}e^{-i\kappa}_{}\left[\begin{array}{c}
A^{}_{L}+B^{}_{L}\\
A^{}_{R}+B^{}_{R}
\end{array}\right]\ ,
\end{align}
where we have used $E=-2J^{}_{0}\cos(\kappa)$. Substituting  into Eq.~(\ref{FJ})   yields
\begin{align}
J^{}_{\rm c}{\cal F}^{-1}_{\lambda}\left[\begin{array}{c}
A^{}_{L}+B^{}_{L}\\
 A^{}_{R}+B^{}_{R}
\end{array}\right]
=2i\frac{J^{}_{0}}{J^{}_{\rm c }}\sin(\kappa)\left[\begin{array}{c}
A^{}_{L} \\
A^{}_{R}
\end{array}\right]+\frac{J^{}_{0}}{J^{}_{\rm c}}e^{-i\kappa}_{}\left[\begin{array}{c}
A^{}_{L}+B^{}_{L}\\
A^{}_{R}+B^{}_{R}
\end{array}\right]\ ,
\end{align}
leading to   the scattering-matrix equation
 \begin{align}
 \left[\begin{array}{c}
 B^{}_{L}\\
 B^{}_{R}
 \end{array}\right]={\cal S}^{}_{\lambda}\left[\begin{array}{c}
 A^{}_{L}\\
 A^{}_{R}
 \end{array}\right]\ ,~~~{\rm with}~~~~
 {\cal S}^{}_{\lambda}\equiv\left[\begin{array}{cc}
 r^{}_{\lambda,l} & t^{ }_{\lambda,r}\\
 t^{}_{\lambda,l} & r^{}_{\lambda,r}
 \end{array}\right]= -1-\frac{2i\sin(\kappa)} {e^{-i\kappa}_{}-(J^{2}_{\rm c}/J^{}_{0}){\cal F}^{-1}_{\lambda}}\ .
 \end{align}
 The unitarity  of the scattering matrix    in both configurations $\lambda=0,1$ is   ensured by the identity  ${\cal F}^{-1}_{}=\left[{\cal F}^{-1}_{}\right]^{\dagger}_{}$.
Finally, using
\begin{align}
\frac{1} {e^{-i\kappa}_{}-(J^{2}_{\rm c}/J^{}_{0}){\cal F}^{-1}_{\lambda}}=\left[\begin{array}{cc}
 \frac{J^{2}_{0}\det[{\cal F}^{}_{\lambda}]e^{-i\kappa}_{}-J^{2}_{\rm c}J^{}_{0}{\cal F}^{}_{\lambda,11}}{J^{2}_{0}\det[{\cal F}^{}_{\lambda}]e^{-2i\kappa}_{}-J^{2}_{\rm c}J^{}_{0}{\rm tr}[{\cal F}^{}_{\lambda}]e^{-i\kappa}_{}+J^{4}_{\rm c}} & -\frac{J^{2}_{\rm c}J^{}_{0}{\cal F}^{}_{\lambda,12}}{J^{2}_{0}\det[{\cal F}^{}_{\lambda}]e^{-2i\kappa}_{}-J^{2}_{\rm c}J^{}_{0}{\rm tr}[{\cal F}^{}_{\lambda}]e^{-i\kappa}_{}+J^{4}_{\rm c}} \\ \\
 -\frac{J^{2}_{\rm c}J^{}_{0}{\cal F}^{}_{\lambda,21}}{J^{2}_{0}\det[{\cal F}^{}_{\lambda}]e^{-2i\kappa}_{}-J^{2}_{\rm c}J^{}_{0}{\rm tr}[{\cal F}^{}_{\lambda}]e^{-i\kappa}_{}+J^{4}_{\rm c}} &  \frac{J^{2}_{0}\det[{\cal F}^{}_{\lambda}]e^{-i\kappa}_{}-J^{2}_{\rm c}J^{}_{0}{\cal F}^{}_{\lambda,22}}{J^{2}_{0}\det[{\cal F}^{}_{\lambda}]e^{-2i\kappa}_{}-J^{2}_{\rm c}J^{}_{0}{\rm tr}[{\cal F}^{}_{\lambda}]e^{-i\kappa}_{}+J^{4}_{\rm c}}
 \end{array}\right]\label{TOTR}\ ,
\end{align}
the  expressions of  the   transmission  and reflection  amplitudes are
\begin{align}
t^{}_{\lambda,l}=&\frac{2iJ^{}_{0}J^{2}_{\rm c} {\cal F}^{}_{\lambda,21}\sin(\kappa)}{J^{2}_{0}\det[{\cal F}^{}_{\lambda}]e^{-2i\kappa}_{}-J^{}_{0}J^{2}_{\rm c}{\rm tr}[{\cal F}^{}_{\lambda}]e^{-i\kappa}_{}+J^{4}_{\rm c}}\ , ~~~~~~~~~
t^{ }_{\lambda,r}=\frac{2iJ^{}_{0}J^{2}_{\rm c} {\cal F}^{}_{\lambda,12}\sin(\kappa)}{J^{2}_{0}\det[{\cal F}^{}_{\lambda}]e^{-2i\kappa}_{}-J^{}_{0}J^{2}_{\rm c}{\rm tr}[{\cal F}^{}_{\lambda}]e^{-i\kappa}_{}+J^{4}_{\rm c}}\ ,  \label{TF}
\end{align}
and
\begin{align}
r^{}_{\lambda,l}=& \frac{J^{2}_{0}\det[{\cal F}^{}_{\lambda}]-J^{}_{0} J^{2}_{\rm c}\left({\cal F}^{}_{\lambda,11}e^{i\kappa}_{}+{\cal F}^{}_{\lambda,22}e^{-i\kappa}_{}\right) +J^{4}_{\rm c}}{J^{2}_{\rm c}J^{}_{0}{\rm tr}[{\cal F}^{}_{\lambda}]e^{-i\kappa}_{}-J^{2}_{0}\det[{\cal F}^{}_{\lambda}]e^{-2i\kappa}_{}-J^{4}_{\rm c}}\ ,~~~
r^{ }_{\lambda,r}=\frac{J^{2}_{0}\det[{\cal F}^{}_{\lambda}]-J^{}_{0} J^{2}_{\rm c}\left({\cal F}^{}_{\lambda,11}e^{-i\kappa}_{}+{\cal F}^{}_{\lambda,22}e^{i\kappa}_{}\right) +J^{4}_{\rm c}}{J^{2}_{\rm c}J^{}_{0}{\rm tr}[{\cal F}^{}_{\lambda}]e^{-i\kappa}_{}-J^{2}_{0}\det[{\cal F}^{}_{\lambda}]e^{-2i\kappa}_{}-J^{4}_{\rm c}}\label{RF}\  .
\end{align}
 Both the reflection and transmission amplitudes depend on the matrix elements of  ${\cal F}^{}_{\lambda}$ and vary with the structural parameters of the chain and the scattering energy $E$.

When the scattering energy $E=-2J^{}_{0}\cos(\kappa)$ lies within the bulk bands, i.e., when $\alpha$  is purely imaginary [see Eq. (\ref{exp})], the function $S^{}_{n}(\alpha)\equiv \sinh(n\alpha)=\sin(n\alpha^{\prime}_{})$, with $\alpha^{\prime}_{}={\rm Im [\alpha^{}_{}]}$. The periodic dependence of $S^{}_{N}(\alpha^{}_{})$ on $N$  and the relation between   $\alpha$ and $E$ [see Eq.~(\ref{exp})], cause the amplitudes to oscillate wildly with the length  $N$  of the chain and to exhibit intricate dependence on the energy  $E$.
Therefore,   it is impractical to try to characterize the topology of the chain in general.
On the other hand,   when $E$ lies  within the band gap,  then  $\exp[\alpha(E)]<-1$,  and therefore
   the  off-diagonal matrix elements ${\cal F}^{}_{\lambda,12}$ and ${\cal F}^{}_{\lambda,21}$  decrease exponentially   as  $N$  increases, and    disappear  for  $N\rightarrow\infty$. We make use of this property below.


\hspace{4mm}
\subsubsection{Scattering matrix at the Fermi energy}
 Fulga {\it et al.}~\cite{Fulga2011,Fulga2012} argued that the topological phase
of a low-dimensional wire belonging to nontrivial topological symmetry classes can be characterized by the sign of a one-sided reflection amplitude at zero energy (i.e., at the Fermi energy).  However,   the reflection amplitudes,  Eq.~(\ref{RF}),  also depend on the coupling to the external leads, $J^{}_{c}$, and on the configuration of the coupling, denoted above by $\lambda$.  {\it Therefore, the sign of the reflection coefficient is not sufficient for identifying the topology.}

When the chain comprises an integral number of unit cells, i.e.,  $\lambda=0$,  the reflection amplitudes  in Eqs.~(\ref{RF}) become [see Eqs.~(\ref{FMS}) and (\ref{detF})]
\begin{align}
r^{}_{0,l}= -\frac{J^{2}_{0}  |W|(|W|S^{}_{N+1}+j^{2}_{2}S^{}_{N})
+J^{}_{0}J^{2}_{\rm c}
[(E|W|S^{}_{N}+2j^{}_{1}j^{}_{2}j^{}_{3}S^{}_{N-1}\cos\Phi-Ej^{2}_{3}S^{}_{N-1})e^{i\kappa}_{}+E|W|S^{}_{N}e^{-i\kappa}_{}]+J^{4}_{\rm c}(|W|S^{}_{N}+j^{2}_{1}S^{}_{N-1})}{J^{2}_{0}  |W|(|W|S^{}_{N+1}+j^{2}_{2}S^{}_{N})e^{-2i\kappa}_{}
+J^{}_{0}J^{2}_{\rm c}
[(E|W|S^{}_{N}+2j^{}_{1}j^{}_{2}j^{}_{3}S^{}_{N-1}\cos\Phi-Ej^{2}_{3}S^{}_{N-1}) +E|W|S^{}_{N} ]e^{-i\kappa}_{}+J^{4}_{\rm c}(|W|S^{}_{N}+j^{2}_{1}S^{}_{N-1})}, \nonumber\\
r^{}_{0,r} =-\frac{J^{2}_{0}  |W|(|W|S^{}_{N+1}+j^{2}_{2}S^{}_{N})
+J^{}_{0}J^{2}_{\rm c}
[(E|W|S^{}_{N}+2j^{}_{1}j^{}_{2}j^{}_{3}S^{}_{N-1}\cos\Phi-Ej^{2}_{3}S^{}_{N-1})e^{-i\kappa}_{}+E|W|S^{}_{N}e^{i\kappa}_{}]+J^{4}_{\rm c}(|W|S^{}_{N}+j^{2}_{1}S^{}_{N-1})}{J^{2}_{0}  |W|(|W|S^{}_{N+1}+j^{2}_{2}S^{}_{N})e^{-2i\kappa}_{}
+J^{}_{0}J^{2}_{\rm c}
[(E|W|S^{}_{N}+2j^{}_{1}j^{}_{2}j^{}_{3}S^{}_{N-1}\cos\Phi-Ej^{2}_{3}S^{}_{N-1}) +E|W|S^{}_{N} ]e^{-i\kappa}_{}+J^{4}_{\rm c}(|W|S^{}_{N}+j^{2}_{1}S^{}_{N-1})}\ .
\end{align}
 At the Fermi energy  i.e.,  for $E=0$,   the wave vector is $\kappa=\pi/2$, and the exponential factors  $\exp[\pm\alpha]$ in Eq.~(\ref{exp}) are given by
$
\exp[ \alpha]= -\frac{j^{2}_{2}}{|j^{}_{1}j^{}_{2}|}$ and $\exp[-\alpha]=-\frac{j^{2}_{1}}{|j^{}_{1}j^{}_{2}|}$ .
Then, one finds  that
\begin{align}
j^{2}_{2}S^{}_{N}+|W|S^{}_{N+1}= \frac{1}{2}
(j^{2}_{1}-j^{2}_{2})e^{-N\alpha}_{}~~{\rm and}~~|W|S^{}_{N}+j^{2}_{1}S^{}_{N-1}= \frac{1}{2}  (j^{2}_{1}-j^{2}_{2} )e^{(N-1)\alpha}_{}\ .
\end{align}
Thus,  the  reflection amplitudes  at the Fermi energy  are
\begin{align}
r^{}_{0,l}=-\frac{J^{2}_{0}  |W|(j^{2}_{1}-j^{2}_{2})e^{-N\alpha}_{}
+4iJ^{}_{0}J^{2}_{\rm c}
 j^{}_{1}j^{}_{2}j^{}_{3}S^{}_{N-1}\cos\Phi  +J^{4}_{\rm c}(j^{2}_{1}-j^{2}_{2} )e^{(N-1)\alpha}_{}}{-J^{2}_{0}  |W|(j^{2}_{1}-j^{2}_{2})e^{-N\alpha}_{}
-4iJ^{}_{0}J^{2}_{\rm c}
 j^{}_{1}j^{}_{2}j^{}_{3}S^{}_{N-1}\cos\Phi +J^{4}_{\rm c}(j^{2}_{1}-j^{2}_{2} )e^{(N-1)\alpha}_{}}\ ,\nonumber\\
 r^{}_{0,r}=-\frac{J^{2}_{0}  |W|(j^{2}_{1}-j^{2}_{2})e^{-N\alpha}_{}
-4iJ^{}_{0}J^{2}_{\rm c}
 j^{}_{1}j^{}_{2}j^{}_{3}S^{}_{N-1}\cos\Phi  +J^{4}_{\rm c}(j^{2}_{1}-j^{2}_{2} )e^{(N-1)\alpha}_{}}{-J^{2}_{0}  |W|(j^{2}_{1}-j^{2}_{2})e^{-N\alpha}_{}
-4iJ^{}_{0}J^{2}_{\rm c}
 j^{}_{1}j^{}_{2}j^{}_{3}S^{}_{N-1}\cos\Phi +J^{4}_{\rm c}(j^{2}_{1}-j^{2}_{2} )e^{(N-1)\alpha}_{}}\ .
\end{align}
Introducing
\begin{align}
\xi= \frac{2j^{}_{1}j^{}_{2}j^{}_{3}\cos\Phi}{j^{2}_{1}
-j^{2}_{2}} ~~{\rm and} ~~~\Lambda=\frac{J^{2}_{\rm c}}{J^{}_{0}}\ ,
\label{chi}
\end{align}
the reflection amplitudes in Eq.~(7) of the main text become
\begin{align}
r^{}_{0,l}=\frac{|W|e^{-N\alpha}_{}
+2i\xi \Lambda S^{}_{N-1}
+\Lambda^{2}_{}e^{ (N-1)\alpha}_{}}{|W|e^{-N\alpha}_{}
+2i\xi \Lambda S^{}_{N-1}
-\Lambda^{2}_{}e^{ (N-1)\alpha}_{}}, ~~~~
r^{}_{0,r}=\frac{|W|e^{-N\alpha}_{}
-2i\xi \Lambda S^{}_{N-1}
+\Lambda^{2}_{}e^{ (N-1)\alpha}_{}}{|W|e^{-N\alpha}_{}
+2i\xi \Lambda S^{}_{N-1}
-\Lambda^{2}_{}e^{ (N-1)\alpha}_{}}\ . \label{R0s}
\end{align}
The  corresponding transmission coefficients in Eq.~(\ref{RF}) are given by
\begin{align}
t^{}_{0,l}=
\frac{2i\Lambda j^{}_{1}e^{-i
(\varphi^{}_{\rm t}+\phi^{}_{1})}_{} }{|W|e^{-N\alpha}_{}
+2i\xi \Lambda S^{}_{N-1}
-\Lambda^{2}_{}e^{ (N-1)\alpha}_{}}\ ,~~~~
t^{}_{0,r}=\frac{2i\Lambda j^{}_{1}e^{i
(\varphi^{}_{\rm t}+\phi^{}_{1})}_{} }{|W|e^{-N\alpha}_{}
+2i\xi \Lambda S^{}_{N-1}
-\Lambda^{2}_{}e^{ (N-1)\alpha}_{}}\ .\label{tlr0}
\end{align}

 Obviously, the sign of a  reflection amplitude can be well-defined only when the amplitude is real.  By Eq.~(\ref{R0s}),   this can happen only  in the nontrivial topological classes, i.e., the ``D'' ($\Phi=\pm\pi/2$) or ``BDI'' ($j^{}_{3}=0$) class in which  $\xi=0$. The reflection amplitudes of both sides  in these  classes are  identical,
 \begin{align}
r^{}_{0,l} =r^{}_{0,r} =\frac{|W|e^{-N\alpha}_{}+\Lambda^{2}_{}e^{(N-1)\alpha}_{}}{|W|e^{-N\alpha}_{}-\Lambda^{2}_{}e^{(N-1)\alpha}_{}}
 \ ,
  \end{align}
and the   sign  of the amplitudes is given  by
\begin{align}
{\rm sign}[r^{}_{0,l}]={\rm sign} [r^{}_{0,r}]=\begin{cases}
1 ~~~~~|j^{}_{1}/j^{}_{2}|>\xi^{}_{N}\\
-1~~~|j^{}_{1}/j^{}_{2}|< \xi^{}_{N}\ ,
\end{cases} \label{nopv}
\end{align}
with $\xi^{}_{N}=\sqrt[N]{|\Lambda/j^{}_{2}|}$.
At the transition point, i.e., $| j^{}_{1}/j^{}_{2}|=\xi^{}_{N}$,   the reflection amplitudes vanish  and the transmission amplitudes are  $|t^{}_{0}|^{2}_{}=|t^{\prime}_{0}|=1$, see Eq.~(\ref{tlr0}).

Compared with the topological criterion derived from the Bloch-band theory,  the sign of the reflection amplitudes is not exclusively determined by the value of the critical factor $|j^{}_{1}/j^{}_{2}|$ but is also related to other parameters of the hybrid system, i.e., $ \Lambda$ and $j^{}_{2}$. The effective parameter range for obtaining a   negative or positive reflection amplitude from either side of a finite chain therefore does not coincide with that of the nontrivial or trivial topological phase as suggested by the Zak phase in Eq.~(\ref{cas}).  For instance,  when  $\Lambda< j^{}_{2}$,   the threshold for attaining a negative reflection amplitude is smaller than the  `standard value'  $1$,    because $\xi^{}_{N}<1$.
An opposite result can be derived in the case $\Lambda>j^{}_{2}$.    Nevertheless, in the thermodynamic limit, the conventional topological analysis can be recovered,   because then  $\lim^{}_{N\rightarrow\infty}\xi^{}_{N}\rightarrow 1$, regardless of the numerical values of  $\Lambda$ and $j^{}_{2}$.   {\it Thus, the Fulga et al. criterion works only for topological phases and only in the thermodynamic limit  $N\rightarrow\infty$}.

 When the unit cell of the rightmost site is imperfect, i.e.,  $\lambda=1$,   the  reflection amplitudes in Eq.~(\ref{RF}) are
 \begin{align}
r^{}_{1,l}= -\frac
{(2j^{}_{1}j^{}_{2}j^{}_{3}S^{}_{N-1}\cos\Phi-
Ej^{2}_{3}S^{}_{N-1}+E|W|S^{}_{N})-i\Lambda
(j^{2}_{1}-j^{2}_{2})S^{}_{N-1}+E\Lambda^{2}_
{}S^{}_{N-1}}{-
(2j^{}_{1}j^{}_{2}j^{}_{3}S^{}_{N-1}\cos\Phi-
Ej^{2}_{3}S^{}_{N-1}+E|W|S^{}_{N})-i\Lambda[
2|W|S^{}_{N}+(j^{2}_{1}+j^{2}_{2})S^{}_{N-1}]+
E\Lambda^{2}_{}S^{}_{N-1}},\nonumber\\
r^{}_{1,r}= -\frac
{(2j^{}_{1}j^{}_{2}j^{}_{3}S^{}_{N-1}\cos\Phi-
Ej^{2}_{3}S^{}_{N-1}+E|W|S^{}_{N})-i\Lambda
(j^{2}_{2}-j^{2}_{1})S^{}_{N-1}+E\Lambda^{2}_
{}S^{}_{N-1}}{-
(2j^{}_{1}j^{}_{2}j^{}_{3}S^{}_{N-1}\cos\Phi-
Ej^{2}_{3}S^{}_{N-1}+E|W|S^{}_{N})-i\Lambda[
2|W|S^{}_{N}+(j^{2}_{1}+j^{2}_{2})S^{}_{N-1}]+
E\Lambda^{2}_{}S^{}_{N-1}}\ ,\label{r1lr}
\end{align}
using the expression for ${\cal F}^{}_{1}$ in Eqs.~(\ref{FMS}).
At the Fermi energy, using  $2|W|S^{}_{N}+(j^{2}_{1}+j^{2}_{2})
S^{}_{N-1}=(j^{2}_{1}-j^{2}_{2})\cosh[(N-1)\alpha]$,
 the reflection amplitudes become
 \begin{align}
 r^{}_{1,l}= \frac{\xi S^{}_{N-1}
 -i\Lambda S^{}_{N-1}
 }{\xi S^{}_{N-1}
 +i\Lambda \cosh[(N-1)\alpha]} ~~~
 {\rm and}~~
 r^{ }_{1,r}=\frac{\xi S^{}_{N-1}
 +i\Lambda S^{}_{N-1}
 }{\xi S^{}_{N-1}
 +i\Lambda \cosh[(N-1)\alpha]}\ .
 \end{align}
The corresponding transmission
amplitudes are
\begin{align}
t^{}_{1,l}=-\frac{i\Lambda  e^{-i\varphi^{}_{\rm t}}_{}}{\xi S^{}_{N-1}
 +i\Lambda \cosh[(N-1)\alpha]} ~~~{\rm and}~~~t^{ }_{1,r}=\frac{i\Lambda  e^{i\varphi^{}_{\rm t}}_{}}{\xi S^{}_{N-1}
 +i\Lambda \cosh[(N-1)\alpha]}\ .
\end{align}
Again,   it is found that real reflection amplitudes can only be obtained in the  ``D'' and ``BDI'' symmetry classes with a zero $\xi$.  However, in this configuration, the   reflection amplitudes of both sides have opposite signs,
\begin{align}
r^{}_{1,l}=-r^{ }_{1,r}=-\frac{\sinh[(N-1)\alpha]}{ \cosh[(N-1)\alpha]}\ .
\end{align}
Interestingly, it is seen that the effect of   coupling to external leads has been removed from    the reflection amplitudes, and
\begin{align}
{\rm sign}[r^{}_{1,l}]=-{\rm sign}[r^{ }_{1,r}]=\begin{cases}
-1 ~~~~~|j^{}_{1}/j^{}_{2}|<1\\
1~~~~~~~~~|j^{}_{1}/j^{}_{2}|>1
\end{cases}\ .
\end{align}
Here, the sign change of the reflection amplitude from the left side is consistent with that of the change in the  Zak phase in Eq.~(\ref{zakx}).   However,  the sign of the reflection amplitude of the other side presents a  different dependence on  $j^{}_{1}/j^{}_{2}$.
 The sign change of the reflection amplitude from either side is accompanied by a unit transmission, i.e.,  $|t^{}_{1,l}|=|t^{ }_{1,r}|=1$,   at the transition point $|j^{}_{1}/j^{}_{2}|=1$.

\begin{figure}
\centering
\includegraphics[width=0.760\textwidth]{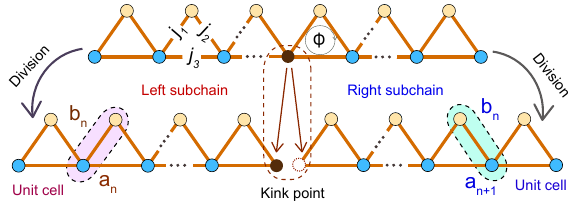}
\caption{(color online)   Schematic diagram of the division of a  finite chain in the $\lambda=1$  configuration into three parts, including the left and right subchains and the kink point of sublattice $a$ located at the middle.  Each subchain is equivalent to a two-sublattice model, but the unit cells of the left and right subchains are different,  consisting of $ \{ b^{}_{n}, a^{}_{n}\}$  on the left subchain, and for by  $\{a^{}_{n+1},b^{}_{n}\}$ on the right one.      }
\label{Figs4}
\end{figure}

The emergence of different reflection amplitudes at the two sides can be illustrated by the particular terminal structure in the  $\lambda=1$ configuration.    Due to the lack of a  $b$-sublattice   at the rightmost site, as shown in Fig.~\ref{Figs4}, the triangle chain in this case cannot be used to simulate a  standard periodic model. Instead, it can be regarded as a hybrid system, composed of two separated two-sublattice subchains and a kink connecting the divided subchains.  Structurally,  the subchain of the left side is isomorphic to a finite chain of   $\lambda=0$ configuration. As a result, a negative reflection amplitude is therefore established,  similar to that of the $\lambda=0$ configuration,   i.e., $|j^{}_{1}/j^{}_{2}|<1$.
Due to the modification of the end structure of the right side,  the subchain of the opposite side exhibits a  different bonding structure. Fortunately,   we find that this subchain can be mapped onto one of the left side by interchanging the roles of the two tunneling amplitudes $j^{}_{1}$ and $j^{}_{2}$. This means that a  unit cell of the subchain is composed by   $ b^{}_{n}$ and $a^{}_{n+1}$,  as seen in Fig.~\ref{Figs4}, and the intra-site and inter-site tunneling amplitudes in this case are represented by   $j^{}_{2}$ and $j^{}_{1}$ respectively. As a result,   the parameter range for a   reflection of negative amplitude on the right side is opposite to that of the other side.


\subsubsection{Scattering matrix in the thermodynamic limit}

According to the bulk-boundary correspondence,  the nontrivial topological phase (as indicated by a topological invariant)  of a chain can be manifested by the presence of an edge state at the Fermi energy. Here we show that the emergence of a  $\pi$-phased reflection amplitude corresponds to the formation of an edge state in an isolated chain in the  {\it thermodynamic}  limit, $N\rightarrow\infty$. In addition,  we demonstrate that this correspondence is not exclusively limited to the nontrivial topological classes, but can also be extended to the trivial-symmetry classes without any restriction on the scattering energy.

  Explicitly, when  the  energy of incoming waves  resonates  with an energy level  of the isolated chain, i.e., $\det[{\cal F}^{}_{\lambda}]=0$,      the reflection amplitudes in Eqs.~(\ref{RF}) are
  \begin{align}
r^{}_{\lambda,l}=
-1-2i\sin(\kappa)\frac{ {\cal F}^{}_{\lambda,11}}{ ({\cal F}^{}_{\lambda,11}+{\cal F}^{}_{\lambda,22})e^{-i\kappa}_{}-\Lambda}
\ ,~~~
r^{ }_{\lambda,r}=-1-2i\sin(\kappa)
\frac{ {\cal F}^{}_{\lambda,22}}
{ ({\cal F}^{}_{\lambda,22}+{\cal F}^{}_{\lambda,11})
e^{-i\kappa}_{}-\Lambda}\ .
\label{rrt}
\end{align}
Clearly,    a  negative reflection amplitude from the left (right) side requires that  ${\cal F}^{}_{\lambda,11/22}=0$. However,   this is incompatible with the above-mentioned condition for the energy spectrum of a {\it finite} isolated chain.    Indeed, since    $\det[{\cal F}^{}_{\lambda}]={\cal F}^{}_{\lambda,11}{\cal F}^{}_{\lambda,22}-{\cal F}^{}_{\lambda,12}{\cal F}^{}_{\lambda, 21}$, the vanishing of a diagonal element implies that $\det[{\cal F}^{}_{\lambda}]=-{\cal F}^{}_{\lambda,12}{\cal F}^{}_{\lambda, 21}$.  However, by Eqs.~(\ref{FMS}),    the off-diagonal matrix elements are generally different from zero, unless the energy is located in the band gap {and the chain length is sufficiently long, i.e.,     $N\rightarrow \infty$.    In other words,  only in the {\it thermodynamic limit}, the appearance of a one-sided edge state in the band gap can be {\it exactly} mapped to the emergence of a $\pi$-phased reflection of that side.

For the configuration of $\lambda=0$,   by the first of Eqs.~(\ref{FMS}),  the expressions  for  ${\cal F}^{}_{0,11}$ and ${\cal F}^{}_{0,22}$ in the thermodynamic limit and with    $\exp[\alpha(E)]<-1$   are
\begin{align}
{\cal F}^{}_{0,11} =-\frac{E|W|e^{\alpha}_{}+2j^{}_{1}j^{}_{2}j^{}_{3}\cos\Phi
-Ej^{2}_{3}}{|W|e^{\alpha}_{}+j^{2}_{1}}\ ,\ \
{\cal F}^{}_{0,22}=-\frac{E|W|e^{\alpha}
_{}}{|W|e^{\alpha}_{}+j^{2}_{1}}\ .
\end{align}
Using the  explicit form of
 $\exp[\alpha]$ in  Eq. (\ref{exp}),
one finds
  \begin{align}
 {\cal F}^{}_{0,11}
 =&\frac{1}{2E}\left[\sqrt{(E+j^{2}_{2}
 -j^{2}_{1})^{2}_{}+
 4E[\xi (j^{2}_{1}-j^{2}_{2})-E^{}_{}(j^{2}_{2}+j^{2}_{3})] }
 -(E^{2}_{}+j^{2}_{2}-j^{2}_{1})\right]\ ,\nonumber\\
  {\cal F}^{}_{0,22}=&\frac{2E|W|e^{\alpha}_{}}{\sqrt{(E+j^{2}_{1}
 -j^{2}_{2})^{2}_{}+
4E[\xi (j^{2}_{1}-j^{2}_{2})-E^{}_{}(j^{2}_{1}+j^{2}_{3})]}
 -(E^{2}_{}+j^{2}_{1}-j^{2}_{2}) }\ ,
 \label{FDS}
 \end{align}
 where   $\xi$ is   given by  Eq.~(\ref{chi}).
As seen in
 the square brackets of   ${\cal F}^{}_{0,11}$ in Eq.~(\ref{FDS}),   a zero ${\cal F}^{}_{0,11}$ requires that
 \begin{align}
  {\rm (i) }~4E[\xi(j^{2}_{1}-j^{2}_{2})-E^{}_{}(j^{2}_{2}+j^{2}_{3})] =0\ , ~~~~~~ {\rm (ii) }~(E^{2}_{}+j^{2}_{2}-j^{2}_{1})\geq0 \ . \label{KH1}
 \end{align}
  (i) implies   two  possibilities of the  energy, i.e., $E=0$ or $ \varepsilon^{}_{2}$, with
  \begin{align}
 \varepsilon^{}_{2}=\frac{\xi(j^{2}_{1}-j^{2}_{2}) }{j^{2}_{2}+j^{2}_{3}}\ .
  \end{align}
   (ii)  restricts the parameter range for attaining a zero ${\cal F}^{}_{0,11}$ at these specific energies.  Note that because the  denominator of ${\cal F}^{}_{0,11}$   also vanishes at $E=0$,  the zero energy   is actually  a {\it wrong} solution to ${\cal F}^{}_{0,11}=0$ unless when $\xi=0$.  To illustrate this, we write  the diagonal matrix element   as
   \begin{align}
   {\cal F}^{}_{0,11}=\frac{1}{2E}\frac{4E[\xi(j^{2}_{1}-j^{2}_{2})-E(j^{2}_{2}+j^{2}_{3})]  }{\sqrt{(E+j^{2}_{2}
 -j^{2}_{1})^{2}_{}+
 4E[\xi (j^{2}_{1}-j^{2}_{2})-E^{}_{}(j^{2}_{2}+j^{2}_{3})] }
 +(E^{2}_{}+j^{2}_{2}-j^{2}_{1})}\ ,
      \end{align}
   which is obviously different from zero   for $E=0$, because  for this energy one has ${\cal F}^{}_{0,11}=-\xi$.
    Thus,
the appearance of     a zero ${\cal F}^{}_{0,11}$ can only  be established if  $E= \varepsilon^{}_{2}$. Combined with the condition  (ii), the   $\pi$-phased reflection of the left side is  given by the second of Eqs.~(11) of the main text, i.e.,
\begin{align}
\varphi^{}_{0,l }=\pi~~~~\forall~\Big\{E=\varepsilon^{}_{2}  \ , ~
j^{2}_{1}<\varepsilon^{2}_{1}+j^{2}_{2}\Big\}\ . \label{rol}
\end{align}

Similarly,  the numerator  of ${\cal F}^{}_{0,22}$ in   Eqs.~(\ref{FDS})   vanishes for zero energy. However,
    it is also necessary to maintain a nonzero denominator at that energy, which implies  $|j^{}_{1}/j^{}_{2}|<1$. The  conditions  for  a  $\pi$-phased reflection  of the right side are
\begin{align}
\varphi^{}_{0,r}= \pi~~~~\forall~\Big\{E=0\ , j^{2}_{1}<j^{2}_{2} \Big\}\ .\label{ror}
 \end{align}
For the topologically-nontrivial symmetry classes, i.e.,  the ``D'' and ``BDI'',  the conditions for a  $\pi$-phased reflection  [see Eq.~(\ref{rol})]  are identical to those of the left side. Out of the  parameter range, i.e., for $j^{2}_{1}>j^{2}_{2}$, then because  ${\cal F}^{}_{0, nn}$ in Eqs.~(\ref{FDS}) diverge and $\kappa=\pi/2$, the reflection phases become zero.  This is consistent with the previous results in  Eq.~(\ref{nopv}) with   $N\rightarrow \infty$.

For the $\lambda=1$ configuration in the thermodynamic limit and with the   energy located in the band gap,  the  diagonal matrix elements of ${\cal F}^{}_{1}$  are
 \begin{align}
 {\cal F}^{}_{1,11} =-\frac{|W|e^{\alpha}_{}+j^{2}_{2}}{E}\  ~~~~~
 {\cal F}^{}_{1,22}=-\frac{|W|e^{\alpha}_{}+j^{2}_{1}}{E}\ ,
 \end{align}
 Using the  form of $\exp[\alpha(E)]$ given  in Eq.~(\ref{exp}),
 it is   found that ${\cal F}^{}_{1,11}={\cal F}^{}_{0,11}$, which
 can be explained by the fact that in both configurations ($\lambda=0$ and $\lambda=1$) the left endpoint has identical surroundings, while
 \begin{align}
 {\cal F}^{}_{1,22}=\frac{1}{2E}\left\{\sqrt{(E+j^{2}_{1}
 -j^{2}_{2})^{2}_{}+
 4E [\xi(j^{2}_{1}-j^{2}_{2})-E^{}_{}(j^{2}_{1}+j^{2}_{3})] }
 -(E^{2}_{}+j^{2}_{1}-j^{2}_{2})\right\}\ .
 \end{align}
Analogously to Eq.~(\ref{KH1}), ${\cal F}^{}_{1,22}$ vanish when
   \begin{align}
   {\rm (i)}.~~ E [\xi (j^{2}_{1}-j^{2}_{2})-E^{}_{}(j^{2}_{1}+j^{2}_{3})]=0 ~~~ {\rm and} ~~~ {\rm (ii)}.~~E^{2}_{}+j^{2}_{1}-j^{2}_{2} \geq 0\ .
   \end{align}
 (i) gives  $E=0$ or $\varepsilon^{}_{1}$  with
 \begin{align}
 \varepsilon^{}_{1}=\frac{\xi(j^{2}_{1}-j^{2}_{2}) }{j^{2}_{1}+j^{2}_{3}}\ .
  \end{align}
 Similar to the discussion above, the zero energy cannot be a solution, since then ${\cal F}^{}_{1,22} \rightarrow-\xi $ except in the topologically nontrivial classes of  $\xi=0$.   The condition for  a $\pi$ reflection of the right side is
 \begin{align}
 \varphi^{}_{1,r}=\pi~~~\forall ~~~ \{E=\varepsilon^{}_{1}\ , \varepsilon^{2}_{1}+j^{2}_{1}>j^{2}_{2} \}\ . \label{r1r}
 \end{align}
Thus the condition for a $\pi$-phased reflection for the right side can be changed by varying the system parameters, in contrast to the condition in Eq.~(\ref{ror}).

Equations~(\ref{rol}), (\ref{ror}), and (\ref{r1r})  present  conditions for   the formation of $\pi$-phased reflections of both $r$ and $r^{\prime}_{}$ in the thermodynamic limit. These results can also be used to estimate the energy and the parameter range available for the edge states of a finite chain in different symmetry classes and for the two endpoints configurations.      For instance,  Fig.~2(a) in the main text shows that the energy  of  the left-side edge state of a  finite chain is not fixed at the Fermi energy for a system in the ``A'' or ``AI'' class.   Instead, the energy of the edge state depends on the system parameters, such as $j^{}_{1}$ and $\Phi$,   and can be described by the expression of the  energy  range, Eq.~(\ref{rol}).  Being consistent with  the  available parameter-range  in Eq.~(\ref{rol}),  it is seen that the edge state    resides in a  larger   range of $j^{}_{1}$,  compared with that  of the (zero-energy) edge state in the ``BDI'' or ``D'' class.


\section{The scattering matrix of a disordered triangle chain}

The total Hamiltonian of a disordered triangle chain coupled to two external leads  is
\begin{align}
H=H^{}_{\Delta} +H^{}_{\rm l}+H_{\rm c}
\end{align}
where
 \begin{align}
 H^{}_{\Delta}= -\sum^{ }_{n }\left(\varepsilon^{}_{n,a} a^{\dagger}_{n}a^{}_{n}+\varepsilon^{}_{n,b} b^{\dagger}_{n} b^{}_{n} \right) -\sum^{ }_{n}\left(j^{}_{n,1}e^{i\phi^{}_{1}}_{}a^{\dagger}_{n}b^{}_{n}+j^{}_{n,2}e^{i\phi^{}_{2}}_{}b^{\dagger}_{n}a^{}_{n+1}+j^{}_{n,3}e^{i\phi^{}_{3}}_{}a^{\dagger}_{n}a^{}_{n+1}+{\rm h.c.}\right)\ .
 \end{align}
The disorder is introduced by allowing different on-site energies on the two sublattices,  $\varepsilon^{}_{n,a}$ and
 $\varepsilon^{}_{n,b}$,   and different,  site-dependent,  tunneling amplitudes. The Hamiltonians of the leads and the couplings of the leads with the chain are the same as the ones presented above. However, the transfer matrix is changed due to the disorder. Below we present the modified   transfer matrix for this case.

 In the numerical simulations, the disorder   is modeled by
 random deviations of the tight-binding parameters away from their values in the ordered chain, i.e.,  $j^{}_{i,n}=j^{}_{i}+\delta j^{}_{i,n}$ ($i=1,2,3$) and $\varepsilon^{}_{a(b), n}=\delta E^{}_{a(b),n}$.  The random numbers  are generated by a software, such as \textit{Mathematica}, and mimic a white noise. For example,  in the specific numerical simulations of Fig.~4 in  the main text, the `noise' of the tunneling amplitude $\delta j^{}_{i,n}$   takes the from of  $0.3\tilde{\nu}$, with  $ \tilde{\nu} $ being a stochastic  number within the range of  $[-0.5,0.5]$.
  When these random numbers are set to zero, our previous analytical results  are recovered.

 The Schr\"{o}dinger equations, at energy $E$,  at the inner sites of the chain are
\begin{align}
E^{}_{} \Psi^{}_{b}(n)=-\varepsilon^{}_{b,n} \Psi^{}_{b}(n) -j^{}_{n,1}
e^{-i\phi^{}_{1}}_{}\Psi^{}_{a}(n)-j^{}_{n,2}e^{i\phi^{}_{2}}_{}\Psi^{}_{a}(n+1)\ ,
\label{ap0}
\end{align}
and
\begin{align}
E^{}_{}\Psi^{}_{a}(n+1)= -\varepsilon^{}_{a,n+1} \Psi^{}_{a}(n+1)
-j^{}_{n+1,1}e^{i\phi^{}_{1}}_{}\Psi^{}_{b}(n+1)-j^{}_{n,2}
e^{-i\phi^{}_{2}}_{}\Psi^{}_{b}(n)
-j^{}_{n+1,3} e^{i\phi^{}_{3}}_{}\Psi^{}_{a}
(n+2)-j^{}_{n,3}e^{-i\phi^{}_{3}}_{}\Psi^{}_{a}
(n) \ .
\label{ap}
\end{align}
Equation (\ref{ap0}) yields
\begin{align}
\Psi^{}_{a} (n+1)= -
\frac{e^{-i\phi^{}_{2}}_{}}{j^{}_{n,2}}
\left[  (E+\varepsilon^{}_{b,n})\Psi^{}_{b}(n)
+j^{}_{n,1} e^{-i\phi^{}_{1}}_{}\Psi^{}_{a}(n)\right]\ ,
\label{am}
\end{align}
 and
\begin{align}
\Psi^{}_{a}(n+2)=
-\frac{e^{-i\phi^{}_{2}}_{}}{j^{}_{n+1,2}}
\left[  (E+\varepsilon^{}_{b,n+1})\Psi^{}_{b}(n+1)
+j^{}_{n+1,1} e^{-i\phi^{}_{1}}_{}\Psi^{}_{a}(n+1)\right]\ ,
\label{am1}
\end{align}
  Inserting Eqs. (\ref{am}) and (\ref{am1})
  into Eq.~(\ref{ap})  gives
\begin{align}
\left[E+\varepsilon^{}_{a,n+1}-j^{}_{n+1,1}e^{-i\Phi}_{}\frac{j^{}_{n+1,3} }{j^{}_{n+1,2}}\right]\Psi^{}_{a}(n+1)= & \left[e^{-i\Phi}_{}(E+\varepsilon^{}_{n+1,b})\frac{j^{}_{n+1,3} }{j^{}_{n+1,2}}-j^{}_{n+1,1}\right]e^{i\phi^{}_{1}}_{}\Psi^{}_{b}(n+1)-j^{}_{n,2}
e^{-i\phi^{}_{2}}_{}\Psi^{}_{b}(n) -j^{}_{n,3}e^{-i\phi^{}_{3}}_{}\Psi^{}_{a}
(n)
\end{align}
 Substituting this expression for  $\Psi^{}_{a}(n+1)$  into  Eq.~(\ref{am}),   one finds
\begin{align}
\Psi^{}_{b}(n+1) = X^{}_{}(n)\Psi^{}_{b}(n) +Y^{}_{}(n)  \Psi^{}_{a}(n)\ ,
\end{align}
with
\begin{align}
X^{}_{}(n)=\frac{e^{-i(\phi^{}_{1}+\phi^{}_{2})}_{}j^{}_{n+1,2}}{e^{-i\Phi}_{}(E+\varepsilon^{}_{n+1,b}) j^{}_{n+1,3} -j^{}_{n+1,1}j^{}_{n+1,2} } \left(j^{}_{n,2}
 - \frac{ E+\varepsilon^{}_{b,n}  }{j^{}_{n,2}} \big[E+\varepsilon^{}_{a,n+1}-j^{}_{n+1,1}e^{-i\Phi}_{}\frac{j^{}_{n+1,3} }{j^{}_{n+1,2}}\big] \right)\ ,
\end{align}
and
\begin{align}
Y^{}_{}(n)=\frac{e^{-i (\phi^{}_{1}+\phi^{}_{3}) }_{}j^{}_{n+1,2}}{e^{-i\Phi}_{}(E+\varepsilon^{}_{n+1,b}) j^{}_{n+1,3} -j^{}_{n+1,1}j^{}_{n+1,2} }\left(j^{}_{n,3} -\frac{  j^{}_{n,1}e^{-i \Phi}_{}  }{j^{}_{n,2}} \big[E+\varepsilon^{}_{a,n+1}-j^{}_{n+1,1}e^{-i\Phi}_{}\frac{j^{}_{n+1,3} }{j^{}_{n+1,2}}\big]\right)\ ,
\end{align}
 where  $\Phi=\phi^{}_{1}+\phi^{}_{2}-\phi^{}_{3}$.
It follows that for sites within the inner part of the chain,
\begin{align}
\left[\begin{array}{c}
\Psi^{}_{b}(n+1)\\
\Psi^{}_{a}(n+1)
\end{array}\right]=\widetilde{\cal T}^{}_{n}\left[\begin{array}{c}
\Psi^{}_{b}(n)\\
\Psi^{}_{a}(n)
\end{array}\right]\ ,
\end{align}
with
\begin{align}
\widetilde{\cal T}^{}_{n}= \left[\begin{array}{cc}
X(n) & Y(n)\\
-\frac{e^{-i\phi^{}_{2}}_{}}{j^{}_{n,2}}(E+\varepsilon^{}_{b,n}) & -\frac{e^{-i(\phi^{}_{2}+\phi^{}_{1})}_{} j^{}_{ n,1}}{j^{}_{n,2}}
\end{array}\right]\ .
\end{align}
 Consequently, the transfer matrix relating the vectors
 $\mathbf{V}^{}_{1}$ and $\mathbf{V}^{}_{N-1}$  is
\begin{align}
\mathbf{V}^{}_{N-1} = \widetilde{\cal  W}^{}_{in } \mathbf{V}^{}_{1}\ ,
\end{align}
 where
\begin{align}
\widetilde{\cal W}^{} _{in}=\widetilde{\cal T}^{}_{N-2}\cdot \cdot \cdot\widetilde{\cal T}^{}_{3}\widetilde{\cal T}^{}_{2}\widetilde{\cal T}^{}_{1}\ .
\end{align}

 It remains to consider the endpoints of the triangle chain.
  When the chain comprises $N$ complete unit cells (the $\lambda=0$ configuration), one finds
 \begin{align}
\Psi^{}_{a}(N)=-\frac{e^{-i\phi^{}_{2}}_{}}{j^{}_{N-1,2}}
\left[(E+\varepsilon^{}_{b,N-1})\Psi^{}_{b}(N-1)
+j^{}_{N-1,1}e^{-i\phi^{}_{1}}_{}\Psi^{}_{a}(N-1)\right]\ ,
 \end{align}
and
\begin{align}
(E+\varepsilon^{}_{a,N})\Psi^{}_{a}(N)=-j^{}_{N,1}e^{i\phi^{}_{1}}_{}
\Psi^{}_{b}(N)-j^{}_{N-1,2}e^{-i\phi^{}_{2}}_{}
\Psi^{}_{b}(N-1) -j^{}_{N-1,3}e^{-i\phi^{}_{3}}_{}
\Psi^{}_{a}(N-1)\ .
\end{align}
 As a result,
\begin{align}
\Psi^{}_{b}(N)= \frac{e^{-i(\phi^{}_{1}+\phi^{}_{2})}_{}}{j^{}_{N,1}} \left[ \frac{(E+\varepsilon^{}_{a,N})(E+\varepsilon_{b, N-1})}{j^{}_{N-1,2}}-j^{}_{N-1,2}\right]\Psi^{}_{b}(N-1)
+\frac{e^{-i(\phi^{}_{3}+\phi^{}_{1})}_{}}{j^{}_{N,1}}\left[\frac{(E+\varepsilon^{}_{a,N})j^{}_{N-1,1}}{j^{}_{N-1,2}}e^{-i\Phi}_{}-j^{}_{N-1,3}\right]\Psi^{}_{a}(N-1)\ ,
\end{align}
 that is,
\begin{align}
\mathbf{V}^{}_{N}=\widetilde{\cal M}\mathbf{V}_{N-1} \label{C-C-1}\ ,
\end{align}
with
\begin{align}
\widetilde{\cal M}=\left[\begin{array}{cc}
\frac{e^{-i(\phi^{}_{1}+\phi^{}_{2})}_{}}{j^{}_{N,1}j^{ }_{N-1,2}} \left( (E+\varepsilon^{}_{a,N})(E+\varepsilon_{b, N-1}) -j^{2}_{N-1,2}\right)&
\frac{e^{-i(\phi^{}_{3}+\phi^{}_{1})}_{}}{j^{}_{N,1}j^{}_{N-1,2}}\left( (E+\varepsilon^{}_{a,N})j^{}_{N-1,1}  e^{-i\Phi}_{}-j^{}_{N-1,3}j^{}_{N-1,2}\right)
\\
-\frac{e^{-i\phi^{}_{2}}_{}}{j^{}_{N-1,2}}
(E+\varepsilon^{}_{b, N-1}) &
-e^{-i(\phi^{}_{2}+\phi^{}_{1})}_{} \frac{j^{}_{N-1,1}}{j^{}_{N-1,2}}
\end{array}\right]\ .
\end{align}
 Thus,  the transfer  matrix equation  between the two outermost ends of the chain in the $\lambda=0$ configuration is
\begin{align}
\mathbf{V}^{}_{N}=\widetilde{\cal M}^{}_{}\widetilde{\cal W}^{}_{in} \mathbf{V}^{}_{1}\ ,
\label{TY}
\end{align}  where  the vector   on  the left side  is
\begin{align}
\left[\begin{array}{c}
\Psi^{}_{b}(1)\\
\Psi^{}_{a}(1)
\end{array}\right]=\widetilde{\cal C}^{}_{L } \left[\begin{array}{c}
\Psi^{}_{a}(1)\\
J^{}_{\rm c} (A^{}_{L}+B^{}_{L})
\end{array}\right]\ ,
\label{CL}
\end{align}
with
\begin{align}
\widetilde{  C}^{}_{L}= \left[\begin{array}{cc}
\frac{j^{}_{1,1}j^{}_{1,3}
e^{-i\Phi}_{}-j^{}_{1,2}(E+\varepsilon_{a,1})}{j^{}_{1,1}j^{}_{1,2}
-(E+\varepsilon^{}_{b,1})j^{}_{1,3}e^{-i\Phi}_{}}e^{-i\phi^{}_{1}}_{} &-\frac{ j^{}_{1,2} }{j^{}_{1,1}j^{}_{1,2}
-(E+\varepsilon^{}_{b,1})j^{}_{1,3}e^{-i\Phi}_{}}e^{-i\phi^{}_{1}}_{}  \\
1&0\end{array}\right]\ .
\label{CL}
\end{align}
 The matrix $\widetilde{\cal C}_{L}$ is obtained by using the Schr\"{o}dinger equation
\begin{align}
(E+\varepsilon_{a,1})\Psi^{}_{a}(1)&=-J^{}_{\rm c}(A^{}_{L}+B^{}_{L})-j^{}_{1,1}e^{i\phi^{}_{1}}_{}\Psi^{}_{b}(1)-j^{}_{1,3}e^{i\phi^{}_{3}}_{}\Psi^{}
_{a}(2)\nonumber\\
&=-J^{}_{\rm c}(A^{}_{L}+B^{}_{L})-j^{}_{1,1}e^{i\phi^{}_{1}}_{}\Psi^{}_{b}(1)-j^{}_{1,3}e^{i\phi^{}_{3}}_{} \left(-\frac{e^{-i\phi^{}_{2}}_{}}{j^{}_{1,2}} [(E
+\varepsilon^{}_{b,1})\Psi^{}_{b}(1)+j^{}_{1,1}
e^{-i\phi^{}_{1}}_{}\Psi^{}_{a}(1)]\right)\ ,
\end{align}
leading to
\begin{align}
\Psi^{}_{b}(1)=\frac{j^{}_{1,1}j^{}_{1,3}
e^{-i\Phi}_{}-j^{}_{1,2}(E+\varepsilon_{a,1})}{j^{}_{1,1}j^{}_{1,2}
-(E+\varepsilon^{}_{b,1})j^{}_{1,3}e^{-i\Phi}_{}}e^{-i\phi^{}_{1}}_{} \Psi^{}_{a}(1)
- \frac{ j^{}_{1,2} }{j^{}_{1,1}j^{}_{1,2}
-(E+\varepsilon^{}_{b,1})j^{}_{1,3}e^{-i\Phi}_{}}e^{-i\phi^{}_{1}}_{} J^{}_{\rm c}
(A^{}_{L}+B^{}_{L})\ .
\end{align}
  On the other edge of the chain
\begin{align}
\left[\begin{array}{c}
\Psi^{}_{b}(N)\\
\Psi^{}_{a}(N)
\end{array}\right]=\widetilde{  C}^{}_{R ,0} \left[\begin{array}{c} J^{}_{\rm c}(A^{}_{R}+B^{}_{R})\\
\Psi^{}_{b}(N)
\end{array}\right]\ ,
\label{CR}
\end{align}
with
\begin{align}
\widetilde{ C}^{}_{R,0 }=\frac{1}{j^{}_{N,1}}\left[\begin{array}{cc}
0&j^{}_{N,1}\\
-e^{i\phi^{}_{1}}_{} &-(E+\varepsilon^{}_{b,N})e^{i\phi^{}_{1}}_{}
\end{array}\right]\ ,
\end{align}
where we have used the   Schr\"{o}dinger  equation
\begin{align}
(E+\varepsilon^{}_{b,N})\Psi^{}_{b}(N)=-j^{}_{1}e^{-i\phi^{}_{1}}_{} \Psi^{}_{a}(N)-J^{}_{\rm c}(A^{}_{R}+B^{}_{R})\ .
\end{align}
  Inserting    Eqs.~(\ref{CL}) and (\ref{CR}) into Eq.~(\ref{TY}) we  find
\begin{align}
\widetilde{  C}^{}_{R ,0} \left[\begin{array}{c} J^{}_{\rm c}(A^{}_{R}+B^{}_{R})\\
\Psi^{}_{b}(N)
\end{array}\right]=\widetilde{  Z}^{}_{0} \left[\begin{array}{c}
\Psi^{}_{a}(1)\\
J^{}_{\rm c} (A^{}_{L}+B^{}_{L})
\end{array}\right]\ ,
\label{OK}
\end{align}
where $\widetilde{  Z}^{}_{ 0}=\widetilde{\cal M}^{}_{}\widetilde{\cal W}^{}_{in}\widetilde{  C}^{}_{L}$.

 When the chain comprises an extra $a$ site (the $\lambda=1$ configuration),  then similar to the derivation of   Eq.~(\ref{C-C-1}), we  find
 \begin{align}
  \left[\begin{array}{c}
 \Psi^{}_{a}(N)\\
 \Psi^{}_{b}(N-1)
 \end{array}\right]=\widetilde{M}^{\prime}_{}  \left[\begin{array}{c}
 \Psi^{}_{b}(N-1)\\
 \Psi^{}_{a}(N-1)
 \end{array}\right]\ ,
 \end{align}
 with
 \begin{align}
\widetilde{M}^{\prime}_{}=\left[\begin{array}{cc}
-\frac{e^{-i\phi^{}_{2}}_{}}{j^{}_{N-1,2}}
(E+\varepsilon^{}_{b, N-1}) &
-e^{-i(\phi^{}_{2}+\phi^{}_{1})}_{} \frac{j^{}_{N-1,1}}{j^{}_{N-1,2}}\\
1&0
\end{array}\right]\ .
 \end{align}
  It then follows that
 \begin{align}
  \left[\begin{array}{c}
 \Psi^{}_{a}(N)\\
 \Psi^{}_{b}(N-1)
 \end{array}\right] =\widetilde{M}^{\prime}_{}{\cal W}^{}_{in} \left[\begin{array}{c}
 \Psi^{}_{b}(1)\\
 \Psi^{}_{a}(1)
 \end{array}\right]\ ,
 \end{align}
 where the leftmost vector   is
 \begin{align}
  \left[\begin{array}{c}
 \Psi^{}_{a}(N)\\
 \Psi^{}_{b}(N-1)
 \end{array}\right]=\widetilde{C}^{ }_{R,1 }  \left[\begin{array}{c}
 J^{}_{\rm c}(A_{R}+B_{R})\\
 \Psi^{}_{a}(N )
 \end{array}\right]\ ,
 \end{align}
where
\begin{align}
\widetilde{C}^{ }_{R,1 }=\left[\begin{array}
{cc}
0&1\\
\frac{e^{i\phi^{}_{2}}_{}j^{}_{N-1,1}}{  (E^{}_{}+\varepsilon^{}_{b,N-1}) j^{}_{N-1,3}e^{i\Phi}_{}-j^{}_{N-1,2}j^{}_{N-1,1}}& \frac{e^{i\phi^{}_{2}}_{} [(E+\varepsilon_{a,N})j^{}_{N-1,1}- j^{}_{N-1,2} j^{}_{N-1,3}e^{i\Phi}_{} ]}{  (E^{}_{}+\varepsilon^{}_{b,N-1}) j^{}_{N-1,3}e^{i\Phi}_{}-j^{}_{N-1,2}j^{}_{N-1,1}}
\end{array}\right]\ .
\label{S33}
\end{align}
 Equation (\ref{S33}) is derived as follows. First, using the Schr\"{o}dinger equations for the two rightmost sites of the chain, i.e.,
 \begin{align}
(E+\varepsilon_{a,N})\Psi^{}_{a}(N)=-j^{}_{N-1,2}e^{-i\phi^{}_{2}}_{}\Psi^{}_{b}(N-1)-J^{}_{\rm c}(A^{}_{R}+B^{}_{R})-j^{}_{N-1,3}e^{-i\phi^{}_{3}}_{}\Psi^{}_{a}(N-1)\
 \end{align}
 and
 \begin{align}
 (E^{}_{}+\varepsilon^{}_{b,N-1} )\Psi^{}_{b}(N-1)= & -j^{}_{N-1,1}
e^{-i\phi^{}_{1}}_{}\Psi^{}_{a}(N-1)-j^{}_{N-1,2}e^{i\phi^{}_{2}}_{}\Psi^{}_{a}(N)\ ,
 \end{align}
 one finds
 \begin{align}
 (E+\varepsilon_{a,N})\Psi^{}_{a}(N)=
 \frac{j^{}_{N-1,2}}{j^{}_{N-1,1}}j^{}_{N-1,3}e^{i\Phi}_{}\Psi^{}_{a}(N)-J^{}_{c}
 (A^{}_{R}+B^{}_{R})+\left[ \frac{ (E^{}_{}+\varepsilon^{}_{b,N-1})}{j^{}_{N-1,1}}j^{}_{N-1,3}e^{i\Phi}_{}-j^{}_{N-1,2}\right]e^{-i\phi^{}_{2}}_{}\Psi^{}_{b}(N-1)\ .
 \end{align}
 As a result,
  \begin{align}
\Psi^{}_{b}(N-1)=\frac{e^{i\phi^{}_{2}}_{}}{  (E^{}_{}+\varepsilon^{}_{b,N-1}) j^{}_{N-1,3}e^{i\Phi}_{}-j^{}_{N-1,2}j^{}_{N-1,1}}\left\{\Big[(E+\varepsilon_{a,N})j^{}_{N-1,1}- j^{}_{N-1,2} j^{}_{N-1,3}e^{i\Phi}_{}\Big]
\Psi^{}_{a}(N)+J^{}_{c}j^{}_{N-1,1}
 (A^{}_{R}+B^{}_{R})\right\}\ .
 \end{align}
 The relation equivalent to Eq. (\ref{OK}) is then
 \begin{align}
 \widetilde{ C}^{ }_{R ,1} \left[\begin{array}{c} J^{}_{\rm c}(A^{}_{R}+B^{}_{R})\\
\Psi^{}_{a}(N)
\end{array}\right]=\widetilde{\cal Z}^{}_{ 1} \left[\begin{array}{c}
\Psi^{}_{a}(1)\\
J^{}_{\rm c} (A^{}_{L}+B^{}_{L})
\end{array}\right] \ ,
\label{OK2}
 \end{align}
 with $\widetilde{ Z}^{}_{ 1}=\widetilde{M}^{\prime}_{}\widetilde{\cal W}^{}_{in}\widetilde{ C}^{}_{L}$.

\vspace{4mm}
 \textbf{Scattering matrix.}  As seen from   Eqs.~(\ref{OK}) and (\ref{OK2}),    for both edge configurations, we find that
\begin{align}
\widetilde{\cal F }^{}_{\lambda}\left[\begin{array}{c}
\Psi^{}_{a}(1)\\
\Psi^{}_{b}(N) \end{array}
\right]=J^{}_{\rm C} \left[\begin{array}{c}
A^{}_{L}+B^{}_{L}\\
A^{}_{R}+B^{}_{R} \end{array}
\right]\ ,
\end{align}
  with
\begin{align}
\widetilde{\cal F}^{}_{\lambda,11}=&\frac{ [\widetilde{C}^{}_{R,\lambda} ]^{}_{21}[\widetilde{Z}^{}_{\lambda}]_{11}-[\widetilde{C}^{}_{R,\lambda}]^{}_{11}[\widetilde{Z}^{}_{\lambda}]_{21} }{[\widetilde{C}^{}_{R,\lambda}]^{}_{11}[\widetilde{Z}^{}_{\lambda}]_{22}-[\widetilde{C}^{}_{R,\lambda}]^{}_{21}[\widetilde{Z}^{}_{\lambda}]_{12} } ~~~~~~~ {\cal F}^{}_{\lambda,12}=\frac{ \det[  \widetilde{C}^{}_{R,\lambda}] }{[\widetilde{C}^{}_{R,\lambda}]^{}_{11}[\widetilde{Z}^{}_{\lambda}]_{22}-[\widetilde{C}^{}_{R,\lambda}]^{}_{21}[\widetilde{Z}^{}_{\lambda}]_{12} }\nonumber\\
 \widetilde{\cal F}^{}_{\lambda,21}=&\frac{ \det[\widetilde{Z}^{}_{\lambda}]}{[\widetilde{C}^{}_{R,\lambda}]^{}_{11}[\widetilde{Z}^{}_{\lambda}]_{22}-[\widetilde{C}^{}_{R,\lambda}]^{}_{21}
 [\widetilde{Z}^{}_{\lambda}]_{12} } ~~~~~~~ \widetilde{\cal F}^{}_{\lambda,22}= \frac{ [\widetilde{C}^{}_{R,\lambda} ]^{}_{22}[\widetilde{Z}^{}_{\lambda}]_{12}-[\widetilde{C}^{}_{R,\lambda}]^{}_{12}[\widetilde{Z}^{}_{\lambda}]_{22} }{[\widetilde{C}^{}_{R,\lambda}]^{}_{11}[\widetilde{Z}^{}_{\lambda}]_{22}-[\widetilde{C}^{}_{R,\lambda}]^{}_{21}[\widetilde{Z}^{}_{\lambda}]_{12} }\label{WIFS}\ .
\end{align}
  The  scattering matrix    of the chain is
\begin{align}
 \left[\begin{array}{c}
 B^{}_{L}\\
 B^{}_{R}
 \end{array}\right]=\widetilde{\cal S}^{}_{\lambda}\left[\begin{array}{c}
 A^{}_{L}\\
 A^{}_{R}
 \end{array}\right]\ ,~~~{\rm with}~~~~
 \widetilde{\cal S}^{}_{\lambda}\equiv\left[\begin{array}{cc}
 r^{}_{\lambda,l} & t^{ }_{\lambda,r}\\
 t^{}_{\lambda,l} & r^{}_{\lambda,r}
 \end{array}\right]= -1-\frac{2i\sin(\kappa)} {e^{-i\kappa}_{}-(J^{2}_{\rm c}/J^{}_{0})\widetilde{\cal F}^{-1}_{\lambda}}\ .
 \end{align}
The reflection and transmission amplitudes of the disordered chain can be derived from the matrix elements of $\widetilde{F}^{}_{\lambda}$  in  Eqs.~(\ref{WIFS}), in analogy with  Eqs.~(\ref{TOTR})-(\ref{RF}). The results are presented in Fig. (4) of the main text.   In addition to the topological symmetry class, e.g., ``BDI'' and ``D'', it illustrates that the presence of nonzero-energy edge states in the band gap of a disordered chain in the lower symmetry classes, e.g., ``A'' and ``AI'',  can also be correlated with the corresponding reflection signals.